\documentclass[12pt]{JHEP} 
\title{Solar Neutrinos Before and After KamLAND}

\author{John N. Bahcall\\
  School of Natural Sciences, Institute for Advanced Study, Princeton,
  NJ 08540\\
    E-mail: \email{jnb@ias.edu}}
\author{M. C. Gonzalez-Garcia \\
  Theory Division, CERN, CH-1211, Geneva 23, Switzerland,\\
  C.N. Yang Institute for Theoretical Physics\\
State University of New York at Stony Brook\\
Stony Brook,NY 11794-3840, USA,\\
and Instituto de F\'\i sica Corpuscular, Universitat de Val\`encia
  -- C.S.I.C. \\ Edificio Institutos de Paterna, Apt 22085, 46071
  Val\`encia, Spain\\
    E-mail: \email{concha@mail.cern.ch}}
\author{Carlos Pe\~na-Garay\\
        School of Natural Sciences, Institute for Advanced Study,
  Princeton, NJ 08540\\
       E-mail: \email{penya@ias.edu}}

\abstract{We use the recently reported KamLAND measurements on oscillations of reactor anti-neutrinos,
together with the data of previously reported solar neutrino experiments, to show that: (1) the total
$^8$B neutrino flux emitted by the Sun is $1.00(1\pm 0.06)$ (1$\sigma$) of the standard solar model
(BP00) predicted flux, (2) the KamLAND measurements reduce the area of the globally allowed oscillation
regions that must be explored in model fitting by six orders of magnitude in the $\Delta m^2-\tan^2
\theta$ plane, (3) LMA is now the unique oscillation solution to a CL of $4.7\sigma$, (4) maximal mixing
is disfavored at $3.1\sigma$, (5) active-sterile admixtures are constrained to $\sin^2\eta\leq 0.13$ at
$1\sigma$, (6) the observed $^8$B flux that is in the form of sterile neutrinos is $0.00^{+0.09}_{-0.00}$
(1$\sigma$), of the standard solar model (BP00) predicted flux, and (7) non-standard solar models that
were invented to avoid completely solar neutrino oscillations are excluded by KamLAND plus solar data  at
$7.9\sigma$. We also refine quantitative predictions for future $^7$Be and $p-p$ solar neutrino
experiments.}


\begin{document}
\input psfig

\section{Introduction}
\label{introduction}

KamLAND is the first experiment to explore with a terrestrial beam the
region of neutrino oscillation parameters that is relevant for solar
neutrino oscillations. This spectacular
achievement~\cite{kamlandannouncement} makes possible new studies of
the physics of neutrinos
~\cite{sterile,friedland,cptconserv,beacom,shrock,aliani-1,grimus,
holanda,bandyo}
and of the solar interior~\cite{cnopaper}. We concentrate here on
refining the allowed regions in neutrino oscillation space and on what
can be learned about the total $^8$B solar neutrino flux, as well as
the sterile neutrino component of the $^8$B neutrino flux. The
procedures we use have been described in
refs.~\cite{sterile,postsno02}.

We assume throughout this paper that CPT is satisfied and that the
anti-neutrino measurements by KamLAND apply directly to the neutrino
section. In fact, the first KamLAND results already show that CPT is
satisfied in the weak sector to a characteristic accuracy of about
$10^{-20}$ GeV~\cite{cptconserv}.

We begin in section~\ref{sec:kamland} by determining the allowed
oscillation regions using the recently released KamLAND data. Our
principal results, which are in good agreement with the analysis of
the KamLAND collaboration~\cite{kamlandannouncement}, are summarized
in figure~\ref{fig:kamlandonly} and in table~\ref{tab:bestfits}.  We
also show that the numerical values of the physical quantities that we
calculate in this paper are insensitive to the details of the analysis
of the KamLAND data (see discussion at the end of
section~\ref{sec:summary}).

We derive in section~\ref{sec:global} a global solution for the neutrino oscillation parameters using,
together with the KamLAND measurements, all the available solar neutrino data from the
chlorine~\cite{chlorine}, gallium~\cite{sage02,gallex,gno}, Super-Kamiokande~\cite{superk}, and
SNO~\cite{snoccnc,snodaynight} experiments. The enormous reduction in the size of the allowed oscillation
regions is evident from a comparison of the `Before' and `After' allowed regions, figure~\ref{fig:before}
and figure~\ref{fig:after}.

We determine in section~\ref{sec:total} the total $^8$B solar neutrino
flux, as well as the uncertainties in the flux that are implied by the
existing experimental data. This flux can be compared directly with
the flux predicted by the standard solar model.

Could there be a large but unobservable flux of $^8$B solar neutrinos
that reaches Earth in the form of sterile (right-handed) neutrinos?
Prior to the announcement of the KamLAND
results~\cite{kamlandannouncement}, the answer to this question was a
resounding ``Yes"\cite{sterile,bargersterile}. At $3\sigma$, the
sterile component of the $^8$B flux could be, prior to KamLAND, as
large as 1.1 times the BP00 predicted flux, which amounts to 50\% of
the total flux for this extreme case. However, the KamLAND measurement
allows us to place a strong upper limit on the flux of sterile $^8$B
neutrinos. We determine in section~\ref{sec:sterile} the best-estimate
sterile $^8$B neutrino flux and the associated uncertainties in the
sterile flux.

We summarize and discuss our main conclusions in
section~\ref{sec:summary}.
\section{KamLAND only}
\label{sec:kamland}

\FIGURE[!ht]{ \centerline{\psfig{figure=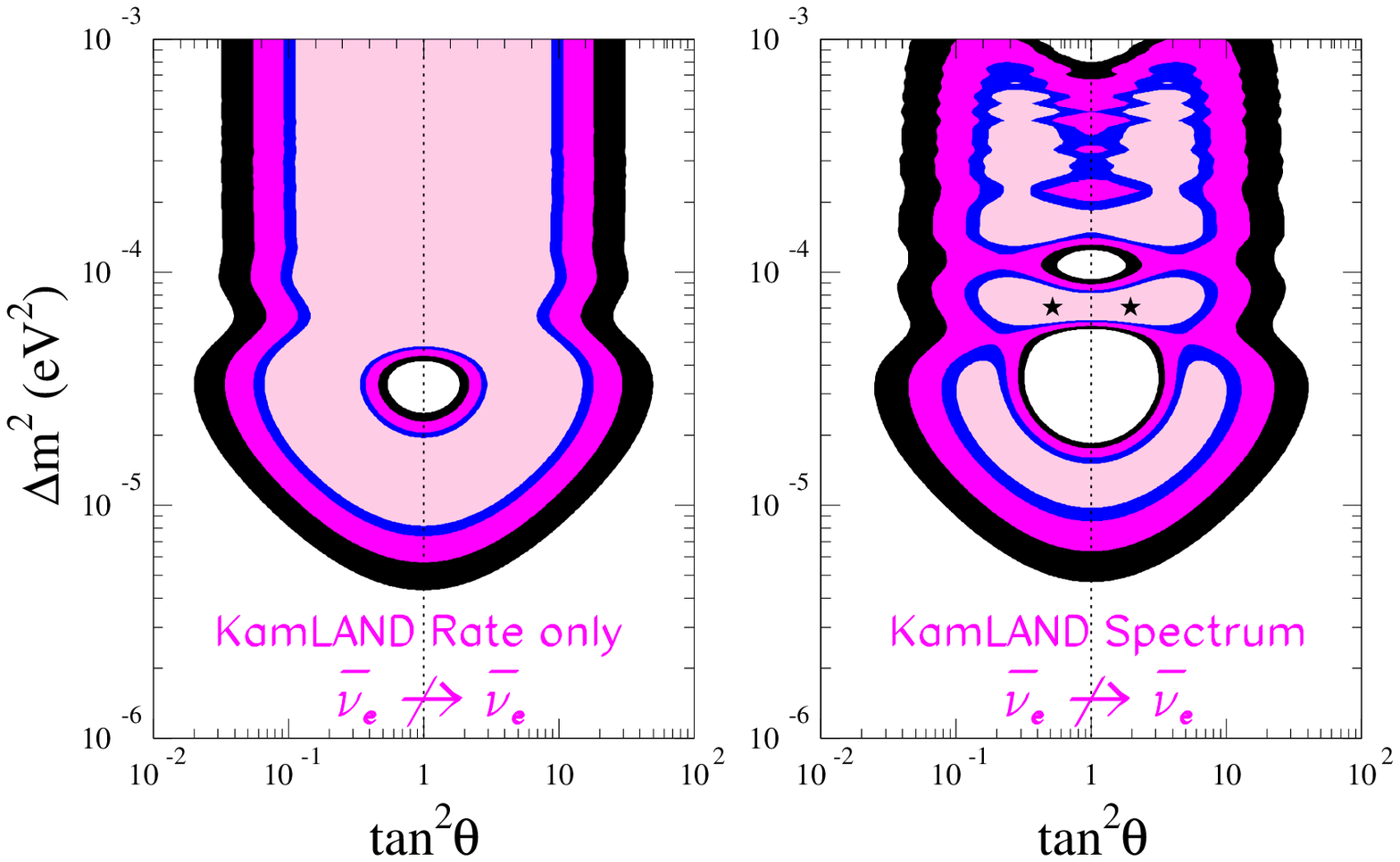,width=5.0in}}
\caption{{\bf Anti-neutrino oscillation parameters regions allowed by
KamLAND.} The figure shows the allowed regions for anti-neutrino
oscillation, $\bar \nu_e ~\not \rightarrow~ \bar \nu_e$, as implied by
the first KamLAND results~\cite{kamlandannouncement}. The contours
shown in the left hand panel were calculated using only the total rate
observed in the KamLAND experiment; the contours in the right hand
panel were calculated using the measured KamLAND energy spectrum. The
contours shown in figure~\ref{fig:kamlandonly} are $90$\%, $95$\%,
$99$\%, and $99.73$\% ($3\sigma$)confidence limits. The dark crosses
mark the best-fit values for the energy spectrum. The non-observation
of anti-neutrino oscillations in the CHOOZ~\cite{chooz} reactor
experiment rules out solutions with $\Delta m^2$ greater than $\sim 1
\times 10^{-3} {\rm eV^2}$ at $3\sigma$ or $\sim 8 \times 10^{-4} {\rm
eV^2}$ at 99\% CL\,.
\label{fig:kamlandonly}}}

Figure~\ref{fig:kamlandonly} shows the allowed regions for
anti-neutrino oscillations that we have determined using the
description of the detector operation and the measured data presented
in the recent KamLAND publication~\cite{kamlandannouncement}. The left
hand panel of figure~\ref{fig:kamlandonly} shows the allowed regions
calculated using only the KamLAND total event rate; the right hand
panel shows the allowed regions calculated using the experimental
energy spectrum, which also contains a rate normalization. In
computing the right hand panel of figure~\ref{fig:kamlandonly}, we
also used the data from the CHOOZ reactor experiment. This experiment
rules out solutions with $\Delta m^2$ greater than $\sim 1 \times
10^{-3} {\rm eV^2}$ at $3\sigma$ or $\sim 8 \times 10^{-4} {\rm eV^2}$
at 99\% CL.

Section~\ref{subsec:kamlandallowed} contains our version (used in the following sections of the paper) of
the science results from the KamLAND measurements and may be of general interest.
Section~\ref{subsec:kamlandanalysis} contains a description of how we
do the data analysis and is rather specialized. We describe in
sec.~\ref{subsec:mattereffects} Earth matter effects in the KamLAND
experiment. The results of sec.~\ref{subsec:mattereffects} show that
matter effects must be included in future precision analyses of the
KamLAND experiment. We have carried out the KamLAND-only and the solar
plus KamLAND analyses described in this paper both with and without
including matter effects for the KamLAND experiment and have verified
that, at the present level of accuracy for KamLAND measurements, the
predictions for future experimental measurements are unaffected by
including matter effects for KamLAND.
Section~\ref{subsec:kamlandanalysis} and
Section~\ref{subsec:mattereffects} are rather technical and can be
skipped by the reader who is primarily interested in the science.

We have verified that different plausible approaches to analyzing the KamLAND data do not lead to
significant differences in the quantities in which we are most interested: the total $^8$B solar neutrino
flux, the sterile component of the $^8$B neutrino flux, the CL at which maximal mixing is excluded, and
the predictions for future solar neutrino experiments.

\subsection{The KamLAND allowed regions}
\label{subsec:kamlandallowed}

As expected, the allowed regions shown in
figure~\ref{fig:kamlandonly} are in agreement with those reported by
the KamLAND collaboration. It was necessary to establish explicitly
this agreement in order to have confidence in the subsequent steps of
our analysis, which involve determining: the globally allowed regions
(section~\ref{subsubsec:globalallowedregions}), predictions for future
solar neutrino experiments
(section~\ref{subsubsec:futureexperiments}), the total $^8$B solar
neutrino flux (section~\ref{sec:total}), an upper limit to the sterile
component of the $^8$B neutrino flux (section~\ref{sec:sterile}), and
an upper limit to the CNO contribution of the solar luminosity
(ref.~\cite{cnopaper}).

The results we obtain for the global solar plus KamLAND
contours of the allowed regions, and all of the quantities derived
from these contours, are essentially independent of the details of the
analysis of the KamLAND results. We have verified this independence by
analyzing the KamLAND measurements in a number of different ways (see
 discussion at the end of section 6).

The most important aspect of figure~\ref{fig:kamlandonly} is the
demonstration by the KamLAND results that anti-neutrinos oscillate
with parameters that are consistent with the LMA solar neutrino
solution~\cite{msw}. We shall exploit this extraordinary achievement
in detail in the following sections.  The symmetry shown in
figure~\ref{fig:kamlandonly} with respect to the line $\theta = \pi/4$
(or $\tan^2 \theta = 1$) is just a reflection of the fact that the
KamLAND experiment is an essentially vacuum experiment (but
cf. sec.~\ref{subsec:mattereffects} and fig.~\ref{fig:matt}) and hence
the survival probabilities depend almost entirely upon $\sin^2
2\theta$.

For the rates-only analysis, the best-fit oscillation parameters form
a degenerate line in the $\Delta m^2$-$\tan^2\theta$ plane. For large
$\Delta m^2$, the line only depends upon $\theta$ and is described by
$\tan^2\theta = 0.36$.

Expressed in terms of the rate expected if there were no flavor changes,
the KamLAND rate is
\begin{equation}
R_{\rm KamLAND} ~=0.611 \pm 0.094\; .
\label{eq:kamlandrate}
\end{equation}
The measured rate is in excellent agreement with the pre-KamLAND
expectations \cite{postsno02}
\begin{equation}
R_{\rm KamLAND, \, expected}  ~= 0.58^{+0.10}_{-0.27}\; ,
\label{eq:kamlandratepred}
\end{equation}
based on purely solar neutrino experiments.

As can be seen in the right hand panel of figure~\ref{fig:kamlandonly} and as is summarized in
table~\ref{tab:kamlandonly}, the KamLAND spectral data lead to three local minima. The allowed region is
separated into `islands.' These islands correspond to oscillations with wavelengths that are
approximately tuned to the average distance between the reactors and the detector, $180$ km. The best-fit
point in the lowest mass island ($\Delta m^2 = 1.5\times 10^{-5}\,{\rm eV^2}$) is near the  first-maximum
in the oscillation probability (minimum in the event rate). The overall best-fit point ($\Delta m^2 =
7.1\times 10^{-5}\,{\rm eV^2}$) lies within the island around the second maximum in the oscillation
probability. The tuning is less accurate for the best-fit point in the highest mass island.

For each of the allowed islands in neutrino oscillation parameter
space that are shown in the right hand panel of
figure~\ref{fig:kamlandonly} (analysis of the energy spectrum),
table~\ref{tab:kamlandonly} gives the best-fit values of $\Delta m^2$
and $\tan^2 \theta$.  The global best-fit point for the KamLAND data
set lies at $\Delta m^2=7.1 \times 10^{-5}$ eV$^2$ and mixing angle
$\tan^2\theta=0.52$. For the same mass difference, maximal mixing is
slightly less favored, $\Delta\chi^2=0.4$.  With the present
statistical accuracy, KamLAND cannot discriminate between large angle
(non-maximal) mixing and maximal mixing.
\TABLE[!t]{ \caption{{\bf KamLAND allowed regions.} The table
presents the best-fit oscillation parameters for the current
3$\sigma$-allowed LMA islands determined by the KamLAND energy
spectrum (see figure~\ref{fig:kamlandonly}). We also give the $\chi^2$
difference between the local minimum in each region and the global
minimum. \label{tab:kamlandonly}}
\begin{tabular}{@{\extracolsep{12pt}}ccc}
\hline\noalign{\smallskip} $\Delta\bar m^2$&$\tan^2\bar\theta$&
$\Delta \chi^2_{min}$ \\
\noalign{\smallskip}\hline\noalign{\smallskip}$7.1\times10^{-5}$ &
$5.2\times10^{-1}$ ($1.9\times10^{0})$ & 0.0 \\ \noalign{\medskip}
$1.7\times10^{-4}$ & $3.5\times10^{-1}$ ($2.8\times10^{0})$ & 1.6\\
\noalign{\medskip} $1.5\times10^{-5}$ & $3.7\times10^{-1}$
($2.7\times10^{0})$ & 3.0 \\ \noalign{\smallskip}\hline
\end{tabular} }

At what level are other, non-LMA oscillation solutions excluded? For
$\Delta m^2 < 10^{-6}{\rm eV^2}$, there is no significant oscillation
probability in the KamLAND experiment. All such low $\Delta m^2$
solutions (like the previously-discussed LOW and Vacuum solar neutrino
oscillation solutions), have $\Delta \chi^2 > 15.9$ and are excluded
by at least $3.6\sigma$ (2 dof).

Comparing the total number of observed and expected events in KamLAND we find that alternative
explanations to the solar neutrino problem that predict no deficit in the KamLAND experiment are now
excluded at 3.6$\sigma$. This applies, for instance, to spin flavor precession, or flavor changing
neutrino interactions which are now ruled out at 3.6$\sigma$ as the dominant mechanism for solar neutrino
flavor conversion. In particular, all `non-standard solar model' explanations of the `solar neutrino
problem' which assume particle physics with $P_{ee}=1$ (no neutrino oscillations) are also excluded by
the KamLAND experiment at $3.6\sigma$.

\subsection{KamLAND analysis}
\label{subsec:kamlandanalysis}

>From the experimental point of view, KamLAND involves different
systematic uncertainties than solar neutrino experiments. From a
theoretical point of view, KamLAND uses anti-neutrinos to perform a
pure disappearance experiment in vacuum.  Solar neutrino studies use
neutrinos to perform disappearance (or appearance) experiments in
matter (or in vacuum). The analysis of the KamLAND data requires a
treatment that is different from, and in many ways more simple than,
the global analysis of solar neutrino data.

Using the data provided in the KamLAND
paper~\cite{kamlandannouncement}, we calculate the positron spectrum
in KamLAND detector with the procedures described in
refs.~\cite{sterile,kamland,prekamlandus,prekamlandthem,robust}. In the
absence of neutrino oscillations, we find (in agreement with
ref.~\cite{kamlandannouncement}) 86.8 expected neutrino events above
2.6 MeV visible energy, for 145.1 days of live time, a 408 ton
fiducial mass with 3.46 $\times 10^{31}$ free target protons,
detection efficiency of 78.3\%, energy resolution
$\sigma(E)/E=7.5\%/\sqrt{E}$, integrated total thermal power flux
during the measurement live time of 254 Joule/${\rm cm^2}$, and
relative fission yields from different fuel components as specified in
ref.~\cite{kamlandannouncement}.  The positron energy spectrum that we
calculate is in excellent agreement with the energy spectrum without
oscillations presented by the KamLAND collaboration.

We perform, following the KamLAND collaboration, two analyses of the
first KamLAND results: 1) including only the total event rate, and 2)
including the total measured energy spectrum. For the rate only
analysis, the $\chi^2$ can be written as ~\cite{kamlandannouncement},
\begin{equation}
\chi^2_{\rm Rate}= \frac{(0.611-R(\theta,\Delta m^2))^2} {\sigma_{\rm
R}^2}\;, \label{eq:krates}
\end{equation}
with $\sigma_{\rm R}^2~=~(0.085)^2~+~(0.041)^2$ and
\begin{equation}
R=1-\sin^2 2\theta \langle\sin^2\left(\frac{\Delta m^2 L}{4 E_\nu}\right) \rangle \; .
\label{eq:Roftheta}
\end{equation}
We denote by $\langle\rangle$ the convolution over the energy spectrum, the interaction cross section,
and the energy resolution function. For simplicity in the notation, we have neglected in writing
eq.~\ref{eq:Roftheta} all matter effects (see sec.~\ref{subsec:mattereffects} for details of the matter
effects).

In the left hand panel of figure~\ref{fig:kamlandonly}, we show the
allowed regions (2 dof) that are obtained from this rate-only
analysis. The vertical asymptotes can be calculated simply from
eq.~(\ref{eq:krates}) since for large $\Delta m^2$ the quantity
$\left<\sin^2\left(\Delta m^2 L/4 E_\nu\right)\right>$ averages to
0.5\,. Thus in the large $\Delta m^2$ limit, the mixing angle boundary
of the allowed region satisfies $\sin^2(2\theta)= -2\sigma_{\rm
R}\sqrt{\Delta\chi^2_{CL}} ~+~0.778$.  Here $\Delta \chi^2$ is fixed
by the CL of interest; $\Delta \chi^2 = 5.99 (11.83)$ for 95\% CL (for
$3\sigma$ CL).

We included the data from the KamLAND energy spectrum in a way that
has become conventional when analyzing low statistics binned data
~\cite{pdg}, assuming that data is Poisson distributed. We define the
$\chi^2$ function as
\begin{eqnarray}
\lefteqn{\chi^2 (\Delta m^2,\theta)=-2{\cal L}=} \nonumber\\ &&{\rm
min}_\alpha\sum_i \left[ 2(\alpha N_i^{\rm th}(\Delta m^2, \theta)-
N_i^{\rm exp})+ 2 N_i^{\rm exp} \ln\left(\frac{N_i^{\rm exp}}{\alpha
N_i^{\rm th} (\Delta m^2, \theta)}\right)\right] +
\frac{(\alpha-1)^2}{\sigma_{\rm sys}^2} \;,\ \ \ \ \ \ \ \
\label{eq:chiklandsp}
\end{eqnarray}
where $\alpha$ is an absolute normalization constant and $\sigma_{\rm
sys}=6.42\%$ represents the systematic uncertainties presented in
table 2 of ref.~\cite{kamlandannouncement}. The KamLAND collaboration
analyzed the event sample by separating the total rate from the
spectrum shape and using a maximum likelihood method.

In the right hand panel of figure~\ref{fig:kamlandonly}, we show the
allowed regions (2 dof) that we obtain from our analysis of the
KamLAND spectrum.

We have performed a series of tests of the dependence of the allowed
solutions on the statistical treatment of the data. For example, we
have performed a standard $\chi^2$ analysis using the first 5 energy
bins above 2.6 MeV visible energy as presented by the KamLAND
collaboration, but with all the events above the fifth bin ($E >
4.725$ MeV) collected into a single bin (or into two bins). We then
assumed Gaussian statistics for all bins. We find that the exact size
and shape of the $3\sigma$ allowed regions of the KamLAND-only
analysis depend mildly on the strategy adopted, i.e., whether we use
eq.~(\ref{eq:chiklandsp}) or the standard $\chi^2$. However, the
differences are unimportant for quantities calculated using the global
solar plus KamLAND allowed regions. We have also studied the effect of
carefully treating the energy dependence of the energy threshold error
by computing, as a function of the oscillation parameters, the
systematic uncertainty for each bin due to the estimated systematic
error for the energy scale (1.91\% at 2.6
MeV\cite{kamlandannouncement}). This refinement has no discernible
effect on any of the quantities we calculate in this paper.

We discuss at the end of section~\ref{sec:summary} the small
quantitative dependence upon analysis strategy of each of the
quantities we calculate in this paper.

\subsection{Matter effects in the KamLAND experiment}
\label{subsec:mattereffects}

It is conventional to ignore matter effects in analyzing the KamLAND results. However, as we shall see
below, matter effects can influence the calculated anti-neutrino event rate by as much as $4$\% if the
mass and mixing angle are near the region of maximum sensitivity. Thus matter effects will have to be
included in future analyses of the KamLAND results when greater statistical and systematic precision are
obtained.

In the results described in this paper, we have included matter
effects by assuming that the anti-neutrinos traverse a constant
density medium ($\rho = 2.7 \,{\rm gm \, cm^{-3}}$, cf.
ref.~\cite{kamland})with the path length appropriate to each reactor.
Since matter effects are small, the assumption of a constant density
is an excellent approximation. Once matter effects are included, the
survival probability for the KamLAND experiment also depends on the
active-sterile admixture.  However, allowing sterile neutrinos reduces
the influence of matter effects with respect to what is calculated for
the pure active case.

To the accuracy that can be discerned by even the most expert eye,
matter effects do not change the contours of figures given in the
present paper.

\FIGURE[!ht]{ \centerline{\psfig{figure=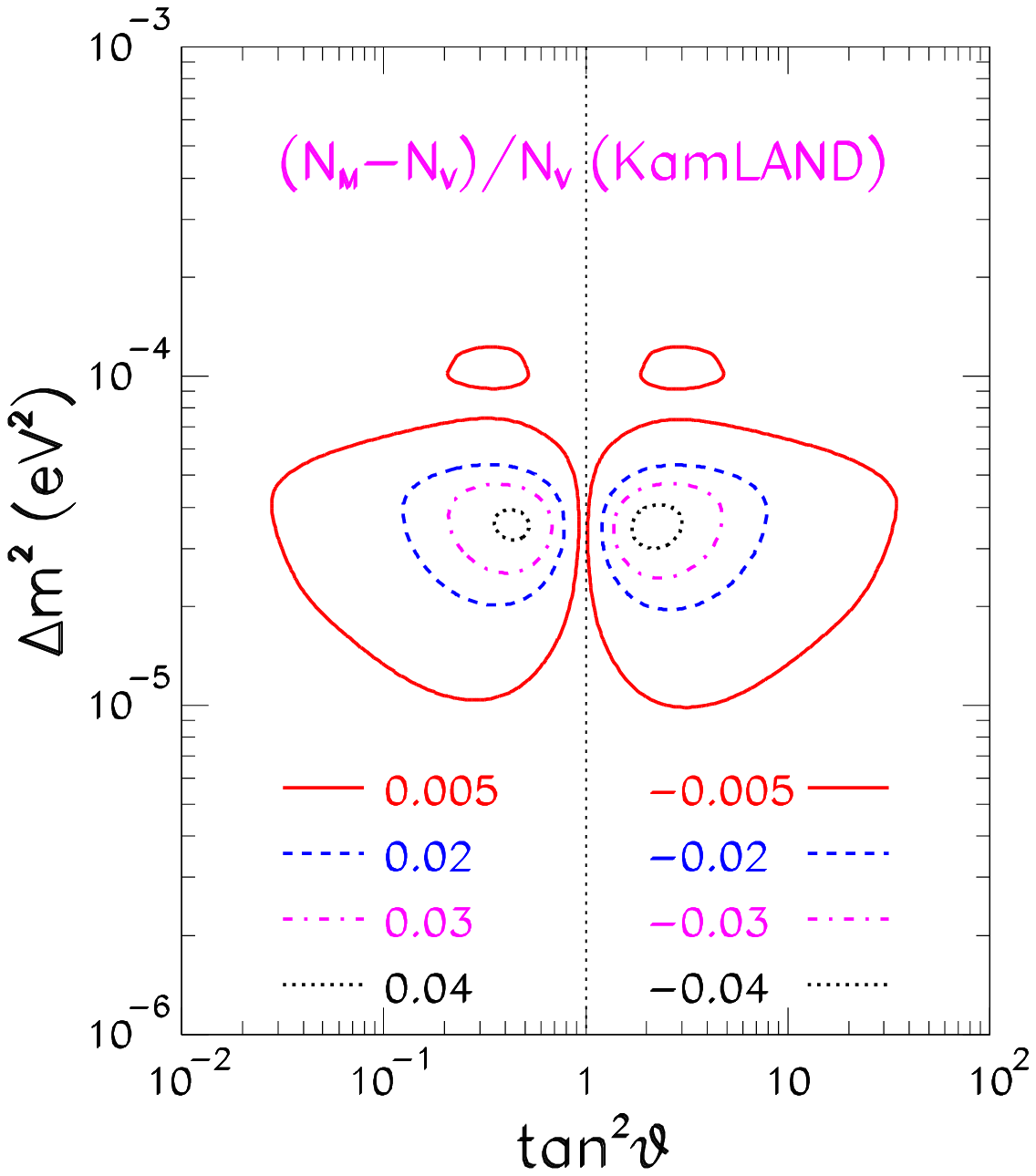,width=3.5in}}
\caption{{\bf Matter effects in the KamLAND reactor experiment.} The
contours depict the fractional difference, defined in
eq.~\ref{eq:fraction}, in the event rate calculated with and without
including matter effects from anti-neutrinos traversing the
Earth. \label{fig:matt}}}

We define, $F({\rm matter~vs~vacuum})$, as the fractional difference
in the event rate for the KamLAND detector calculated with and without
including matter effects in the Earth. Thus
\begin{equation}
F({\rm matter~vs~vacuum}) ~=~ \frac{R_{\rm matter} - R_{\rm
 vacuum}}{R_{\rm vacuum}} \, .  \label{eq:fraction}
\end{equation}

Figure~\ref{fig:matt} shows the contours of $F({\rm
matter~vs~vacuum})$ in the $\Delta m^2$-$\tan^2\theta$ plane for
active neutrinos. The maximum value of $|F({\rm matter~vs~vacuum})|$
corresponds to -4.2\% (+4.3\%) at $\Delta m^2 ~=~3.55\times10^{-5}
{\rm ~eV^2}$ and $\tan^2 \theta = 0.43 (2.18)$. Within the $1\sigma
(3\sigma)$ allowed region of the analysis of the KamLAND energy
spectrum (cf. the right hand panel of fig.~\ref{fig:kamlandonly}), the
maximum value of $|F({\rm matter~vs~vacuum})|$ corresponds to $0.9$\%
($3.9$\%). The maximum change in $\chi^2$ due to including matter
effects in the KamLAND-only analysis is 0.2 (1.5) at $1\sigma
(3\sigma)$. At the best-fit point for the analysis of the KamLAND
energy spectrum, matter effects change the value of $\chi^2$ by only
0.1\%.

We anticipate the results given latter in sec.~\ref{sec:global} (see especially fig.~\ref{fig:after}) by
noting that the matter effects in the KamLAND experiment are somewhat less important within the globally
allowed solar plus KamLAND oscillation regions. For the global solution, $|F({\rm matter~vs~vacuum})|$
varies between $0.9$\% ($2.7$\%) at $1\sigma (3\sigma)$. The maximum change in $\chi^2$ in the global
analysis due to matter effects is 0.2 (1.0).

For simplicity, we have described in the rest of this paper the KamLAND experiment as if matter effects
could be completely ignored.

\section{Global solutions: solar plus KamLAND}
\label{sec:global}

We present and discuss in section~\ref{subsec:globalscientificresults}
the allowed regions in neutrino oscillation parameter space that are
permitted at $3\sigma$ by the combined solar and KamLAND experimental
data (see especially figure~\ref{fig:after}). We also present in this
section (see especially table~\ref{tab:predictions}) the predictions
of future solar neutrino observables that are calculated using the
current global oscillation solutions.

In order to isolate the crucial measurements that have identified LMA
as the correct description of solar neutrino oscillations, we present
in section~\ref{subsec:ratesonly} the results of an analysis of just
the rates of the solar neutrino experiments. We show in this section
that, if we did not have the essentially undistorted electron recoil
energy spectrum measured by Super-Kamiokande and confirmed by SNO,
vacuum oscillations would be the preferred solution prior to the
KamLAND measurements.

We include two sections that are primarily intended for aficionados of neutrino oscillation analyses. We
outline briefly in section~\ref{subsec:kamlandplussolar} how we have combined the solar and the KamLAND
analyses. We have already described in section~\ref{subsec:kamlandanalysis} how we have analyzed the
KamLAND measurements. In section~\ref{subsubsec:solaranalysis}, we summarize our treatment of the solar
data.  We describe in section~\ref{subsubsec:sterile} our treatment of the sterile component of the solar
neutrino flux.

\subsection{Scientific results}
\label{subsec:globalscientificresults}

We describe in section~\ref{subsubsec:globalallowedregions} the
allowed regions in neutrino oscillation space that are determined by
the solar and KamLAND data and in
section~\ref{subsubsec:futureexperiments} we present predictions for
future solar neutrino experiments of the presently allowed global
oscillation solutions.

\subsubsection{Allowed regions}
\label{subsubsec:globalallowedregions}

Figure~\ref{fig:before} presents the `Before' figure, the allowed
neutrino oscillation regions for all the available solar neutrino data
prior to the announcement of the first KamLAND results. We show in
figure~\ref{fig:before} the results of our analysis for this paper of
the pre-KamLAND solar neutrino data. We include here for the first
time the recently reported GNO measurements~\cite{gno}; this inclusion
does not affect figure~\ref{fig:before} significantly.  Our
calculational procedures, which are described in
ref.~\cite{postsno02}, contain some refinements not taken account of
in other analyses.  However, the overall results for the pre-KamLAND
allowed regions are very similar in most treatments,
cf.~\cite{bargernc}.

\FIGURE[!ht]{
\centerline{\psfig{figure=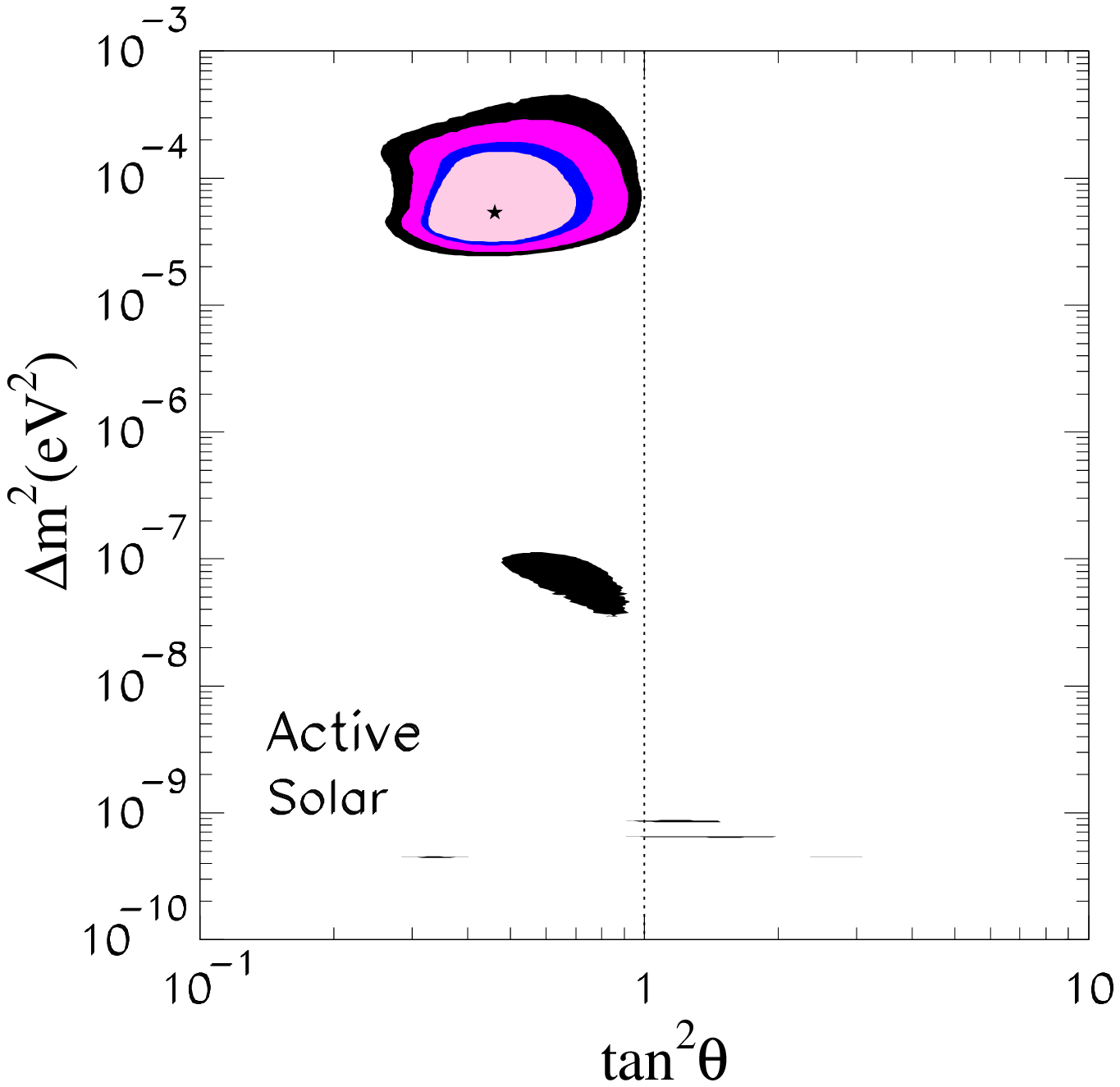,width=3.5in}}
\caption{{\bf Before: Solar neutrino oscillations before KamLAND.} The
contours shown in figure~\ref{fig:before} are $90$\%, $95$\%, $99$\%,
and $99.73$\% ($3\sigma$)confidence limits. The global best-fit points
are marked by a star.The input data used in constructing the figure
include all the available solar neutrino
measurements~\cite{chlorine,snoccnc,snodaynight,sage02,gallex,gno,superk}
and the neutrino fluxes and uncertainties predicted by the BP00 solar
model~\cite{bp00} for all but the $^8$B flux. The $^8$B neutrino flux
is treated as a free parameter.\label{fig:before}}}

Figure~\ref{fig:after} displays the `After' figure, the much reduced
allowed regions that are determined by including the recently
announced KamLAND measurements\cite{kamlandannouncement} together with
the solar neutrino data.

\FIGURE[!ht]{
\centerline{\psfig{figure=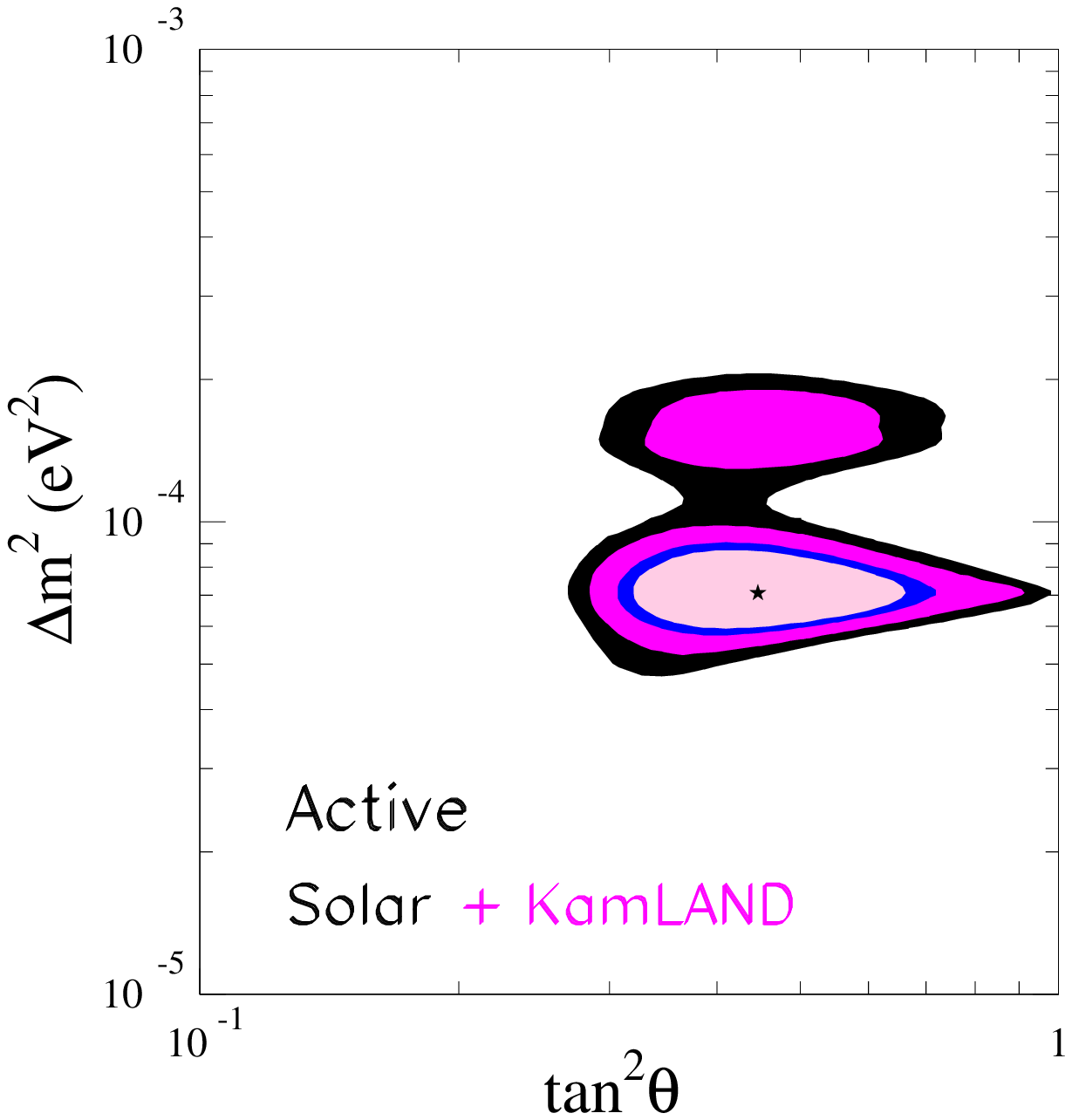,width=3.5in}}
\caption{{\bf After: Solar neutrino oscillations after KamLAND.} The
allowed contours shown in figure~\ref{fig:before} include the recent
KamLAND measurements and all the available solar neutrino
data~\cite{chlorine,sage02,gallex,gno,superk,snoccnc,snodaynight}. The
contour limits are the same as in figure~\ref{fig:kamlandonly} and
figure~\ref{fig:before}. \label{fig:after}}}

Comparing figure~\ref{fig:before} with figure~\ref{fig:after}, we see
that KamLAND has shrunk enormously the allowed parameter space that is
accessible to theoretical models, by five orders of magnitude in
$\Delta m^2$ and one order of magnitude in $\tan^2 \theta$.  The LMA
region in parameter space is the only remaining allowed solution. The
best-fit LOW solution is excluded at $4.8\sigma$ and vacuum solutions
are excluded at $4.9\sigma$. The once `favored' SMA solution is now
excluded at $6.1\sigma$.

The allowed range for the mixing angle within the 1$\sigma$
(3$\sigma$) region shown in the `After' figure~\ref{fig:after} is
\begin{equation}
 0.36 (\,0.27) < \tan^2 \theta < 0.58\, (0.94)\,.
\label{eq:newtanlimitlma}
\end{equation}
For the pre-KamLAND analysis, figure~\ref{fig:before}, the allowed $3\sigma$ domain for the mixing angle
was $0.26< \tan^2 \theta < 0.98$. Therefore, the allowed range of mixing angles has only been marginally
improved by the inclusion of the KamLAND data.

At what confidence level is maximal mixing excluded? There are no solutions with $\tan^2 \theta = 1$ at
the $3.1\sigma$ CL. For the pre-KamLAND analysis, figure~\ref{fig:before}, there were no solutions with
maximal mixing within the LMA region at $3.1\sigma$. However, maximal mixing was still allowed at
$2.8\sigma$ within the vacuum
solution regions, which is now excluded.

Figure~\ref{fig:pee} shows how solar neutrino experiments break the symmetry of the survival probability,
$P_{\rm ee}\left(E\right)$, about a mixing angle of $\theta = \pi/2$ that exists for vacuum oscillation
experiments. Vacuum oscillation survival probabilities are identical for two mixing angles $\theta_1$ and
$\theta_2$, with $\theta_2 = \pi/2 ~-~ \theta_1$ and a fixed value for $\Delta m^2$. The left hand panel
shows the vacuum survival probabilities for a source-detector separation of 180 km, similar to the actual
experimental situation for KamLAND. The right hand panels displays the daytime survival probabilities for
the two LMA solar neutrino oscillation solutions. Matter effects in the Sun are responsible for the
different LMA survival probabilities.

Which solar neutrino experiments choose the lower probabilities (the
smaller value for $\tan^2\theta$)?  Originally, the chlorine
experiment, in conjunction with the standard solar model predictions,
indicated strongly that the smaller probabilities were correct. The
measured event rate in the chlorine experiment, $\sim 2.6$
SNU~\cite{chlorine}, was closer to $0.35$ of the standard solar model
prediction than to $0.65$ of the solar model prediction ($\sim 8$
SNU)~\cite{bp00}. This result was confirmed by the comparison of the
Super-Kamiokande $\nu-e$ scattering measurements with the SNO CC
measurement~\cite{sno01}. The comparison within the same experiment of
the SNO CC and NC rates selects unambiguously the curve with the lower
survival probabilities~\cite{snoccnc}.

\FIGURE[!ht]{ \centerline{\psfig{figure=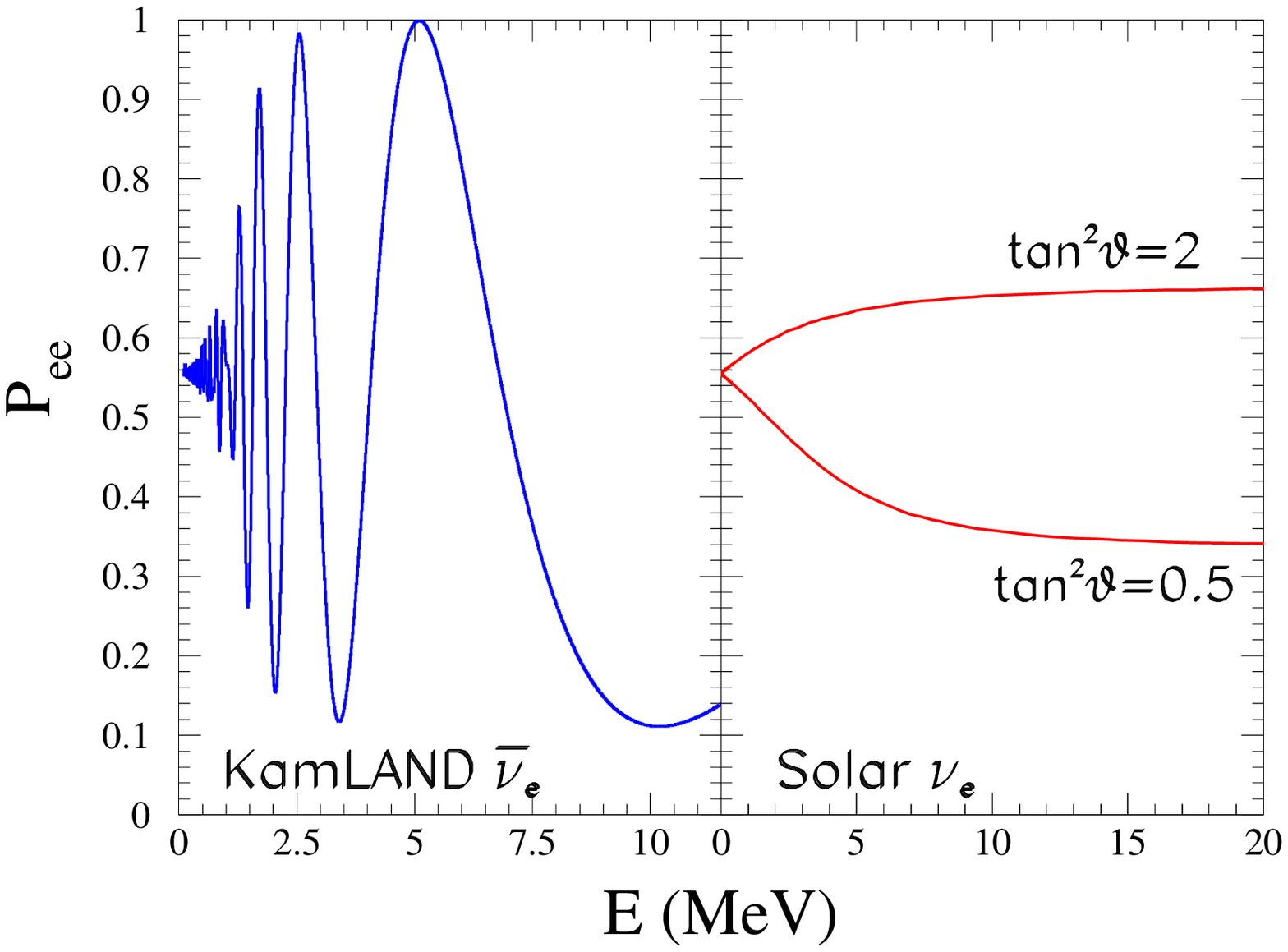,width=3.5 in}}
\caption{{\bf Breaking the degeneracy: $P_{\rm ee}$ vs $E$.}
Figure~\ref{fig:pee} shows the survival probability, $P_{\rm ee}$, as
a function of neutrino energy, $E$, for two mixing angles related by
$\theta_2 = \pi/2 ~-~ \theta_1$ ($\theta_1 = 35.3^0, \tan^2\theta_1 =
0.5, \tan^2 \theta_2 = 2.0$), with $\Delta m^2=7\times 10^{-5} {\rm
~eV^2}$. The left hand panel shows the survival probability (averaged
over $0.1$ MeV energy bins) in vacuum for a source-detector separation
of $180$ km. The vacuum survival probabilities are identical for
$\theta_1$ and $\theta_2$. The right hand panel shows the daytime LMA
oscillation solutions for solar neutrinos.  Matter effects in the Sun
cause the differences between the two solar
solutions. \label{fig:pee}}}

Finally, we have evaluated the present status of `non-standard' solar model explanations of the solar
plus KamLAND data under the hypothesis of standard particle physics. We perform a fit to the total
measured rates in the chlorine~\cite{chlorine}, gallium~\cite{sage02,gallex,gno},
Super-Kamiokande~\cite{superk}, SNO CC~\cite{snoccnc}, and SNO NC~\cite{snoccnc} experiments, allowing
the dominant $p-p$, $^8$B, $^7$B and CNO fluxes to vary freely but imposing the luminosity constraint
(see \cite{luminosity} and references therein for details). We find that $\chi^2_{\rm min, solar}=53.9$
for $2$ dof (5 rates - 4 free fluxes + 1 constraint), which implies that non-standard solar model
explanations of the solar data can be excluded at $7\sigma$. Adding to the analysis the evidence of
antineutrino disappearance observed in the KamLAND experiment, we find that non-standard solar models
invented to avoid solar neutrino oscillations are excluded by solar plus KamLAND experiments at
$7.9\sigma$.

\subsubsection{Predictions for future experiments}
 \label{subsubsec:futureexperiments}

Table~\ref{tab:predictions}
presents the best-fit predictions and the
expected ranges for some important solar neutrino
observables. The results summarized in table~\ref{tab:predictions}
correspond to the globally allowed oscillation regions shown in
figure~\ref{fig:after}.
We consider predictions for SNO,
for a large SK-like neutrino-electron scattering experiment, and for a
neutrino-electron scattering detector, BOREXINO~\cite{borexino} or
KamLAND~\cite{kamland}, that observes the $^7$Be solar neutrinos.
We also include predictions for a generic $p-p$ neutrino detector. The
notation used here is the same as in ref.~\cite{postsno02}.

\TABLE[h!t]{\caption{\label{tab:predictions}{\bf Post-KamLAND
Predictions.} This table presents for future solar neutrino
observables the best-fit predictions, and the $1\sigma$ and $3\sigma$
ranges, that were obtained from the global analysis shown in
figure~\ref{fig:after}.}
\begin{tabular}{lcc}
\hline \noalign{\smallskip} Observable& b.f. $\pm 1\sigma$ &b.f $\pm 3\sigma$ \\
\noalign{\smallskip}\hline \noalign{\smallskip} A$_{\rm N-D}$ (SNO CC) (\%) & $3.3^{+0.7}_{-0.6}$
& $3.3^{+5.7}_{-3.1}$\\
\noalign{\smallskip} \noalign{\smallskip} A$_{\rm N-D}$ (SK ES) (\%) & $1.9\pm 0.4$
& $1.9^{+2.5}_{-1.8}$\\
\noalign{\smallskip} \noalign{\smallskip}
$[$R
($^7$Be)$]$ $\nu-e$ scattering & $0.64\pm 0.02$ &
$0.64^{+0.08}_{-0.04}$ \\ \noalign{\smallskip} \noalign{\smallskip}
A$_{\rm N-D}$ ($^7$Be) (\%) & $0.0^{+0.0}_{-0.0}$ &
$0.0^{+0.1}_{-0.0}$ \\ \noalign{\smallskip} \noalign{\smallskip}$[$R
($^8$B)$]$ $\nu-e$ scattering & &\\ ($T_{\rm th}=3.5$ MeV) & $0.46\pm
0.03$ & $0.46\pm 0.04$ \\ ($T_{\rm th}=5$ MeV) & $0.45\pm 0.03$ &
$0.45\pm 0.04$ \\ \noalign{\smallskip} \noalign{\smallskip} $[p$-$p]$
$\nu-e$ scattering & & \\ ($T_{\rm th}=100$ keV) & $0.697\pm 0.023$ &
$0.697^{+0.064}_{-0.042}$\\ ($T_{\rm th}=50$ keV) & $0.693\pm 0.024$ &
$0.693^{+0.064}_{-0.043}$\\ \noalign{\smallskip} \hline
\end{tabular}
}
The day-night asymmetry $A_{\rm N-D}$ is defined in terms of the Day
and the Night event rates by the expression
\begin{equation}
A_{\rm N-D} ~=~2{\rm {[Night - Day]} \over {[Night +
Day]}}~. \label{eq:daynightdefn}
\end{equation}

The prediction for the day-night asymmetry in the SNO CC measurement
is one of the most interesting new results to come from the much
smaller, post-KamLAND allowed regions (see discussion of
earth matter effects in the LMA regime
in refs.~\cite{obslma,10comm,zenith}).

Within the unique 1$\sigma$ region (2 dof), the prediction is
\begin{equation}
A_{\rm N-D}({\rm SNO~CC}) ~=~ 3.3^{+0.7}_{-0.6} \% \,.
 \label{eq:CCdaynight}
\end{equation}
Within the $3\sigma$ region (2 dof), $A_{\rm N-D}({\rm SNO ~CC})~=~3.3^{+5.7}_{-3.1} \% \,.$

An accurate day-night asymmetry measurement in the SNO CC mode could, in principle, help to discriminate
between larger and smaller values of $\Delta m^2$. For values of $\Delta m^2 < 10^{-4}{\rm eV^2}$ within
the currently allowed $3\sigma$ (2 dof) region,
\begin{equation}
A_{\rm N-D}({\rm SNO~CC}) ~=~ 3.3^{+5.7}_{-1.6} \% \,, ~(\Delta m^2 < 10^{-4}\,{eV^2}) \,.
\label{eq:CCdaynight_b}
\end{equation}
For values of $\Delta m^2 \ge 10^{-4}{\rm eV^2}$, the day-night asymmetry
is small,
\begin{equation}
A_{\rm N-D}({\rm SNO~CC}) ~=~ 0.6^{+1.1}_{-0.4} \% \,, ~(\Delta m^2 \ge 10^{-4}\,{eV^2}) \,.
\label{eq:CCdaynight_a}
\end{equation}
Unfortunately, it will be very difficult to reach the required experimental precision.

 In table~\ref{tab:predictions} we also present the predicted day-night asymmetry at a
SK-like neutrino-electron scattering detector with a threshold $E_{\rm th}=5$ MeV. For ES the day-night
effect is decreased relative to pure CC interactions by the contribution of the neutral currents.

The prediction for the reduced $^7$Be $\nu-e$ scattering rate,
\begin{equation}
[{\rm ^7Be}] ~\equiv~ \frac{\rm Observed ~\nu-e ~scattering~ rate}{\rm
BP00~ predicted ~rate}~,
\label{eq:be7reducedrate}
\end{equation}
is remarkably precise and remarkably stable. Within the $1\sigma$ allowed region, the pre-KamLAND
prediction was $[{\rm ^7Be}]~= 0.64 \pm 0.03, ~1\sigma$~\cite{postsno02}.  The post-KamLAND prediction
for the currently allowed $1\sigma$ region is

\begin{equation}
\left[{\rm ^7Be}\right]_{\rm post-KamLAND}~= 0.64\pm 0.02 \,.
 \label{eq:be7prediction}
\end{equation}
 The predicted event rate for the $^7$Be rate experiment has a
precision of $\pm 3$\% in units of the rate calculated with the BP00
flux and assuming no neutrino oscillations.

Now that LMA is the only oscillation solution allowed at $3\sigma$, the predicted $^7$Be day-night
asymmetry averaged over a year is too small to be observed.  Within the current 3$\sigma$ allowed
oscillation region, the limit on the predicted values of $A_{\rm N-D}$ is
\begin{equation}
\left|A_{\rm N-D}({\rm ^7Be})\right| ~<~ 0.001.
 \label{eq:7Bedaynightlimit}
\end{equation}

We also include in table~\ref{tab:predictions} predictions for the
rate at which $^8$B neutrinos will be scattered by electrons resulting
in recoil electrons in BOREXINO or KamLAND above two different
electron recoil kinetic-energy energy thresholds, $T_{\rm th}=3.5$ MeV
and $T_{\rm th}=5$ MeV. The predictions are given in terms of the
reduced event rate for $^8$B neutrinos, [R($^8$B)], which is defined
analogously to [$^7$Be] [see eq.~(\ref{eq:be7reducedrate})] relative
to the scattering rate predicted for the standard solar model flux
with no oscillations. The predicted ranges of [R($^8$B)] are obtained
from the allowed 1$\sigma$ or 3$\sigma$ regions in the three parameter
space of $\Delta m^2, \tan^2\theta$, and $f_{\rm B,total}$. Not
surprisingly, the predicted $^8$B neutrino-electron scattering rates
are essentially identical to the neutrino-electron scattering rates
measured by the SNO~\cite{snoccnc} and Super-Kamiokande~\cite{superk}
experiments, which are $0.47 \pm 0.05$ and $0.465 \pm 0.015$,
respectively.  For plausible assumptions about how the BOREXINO
detector will operate, one expects~\cite{postsno02} $\sim 175$ events
above a $3.5$ MeV recoil electron energy threshold and $\sim 110$
events per year above a $5$ MeV threshold.

In addition, we have evaluated the predictions for a generic $p-p$
neutrino-electron scattering detector.  The reduced rate for this
detector is defined by the relation
\begin{equation}
[p-p] ~\equiv~ \frac{\rm Observed ~\nu-e ~scattering~ rate}{\rm BP00~
predicted ~rate}~.
\label{eq:ppreducedrate}
\end{equation}
We present the predicted rate for two plausible kinetic energy
thresholds, $100$ keV and $50$ keV. The predicted rate is precise and
robust. The precision of the prediction is $\pm 3$\%; the best-fit
predictions have changed by less than $1$\% from the pre-KamLAND
predictions~\cite{postsno02}.

The principal differences between the post-KamLAND predictions
summarized in table~\ref{tab:predictions} and the pre-KamLAND
predictions given in table~2 of ref.~\cite{postsno02} are a
consequence of the disappearance of the LOW and the vacuum
solutions. As a consequence, no significant day-night asymmetry nor
seasonal variations are expected for $^7$Be neutrino measurements by
BOREXINO or KamLAND. For all other observables, the main effect is the
reduction of the 1$\sigma$ LMA ranges. Within the 3$\sigma$ LMA
ranges, the changes are generally insignificant.

\subsection{Rates only analysis}
\label{subsec:ratesonly}

It is instructive to carry out a global analysis using only the total
measured rates in the chlorine~\cite{chlorine},
gallium~\cite{sage02,gallex,gno}, Super-Kamiokande~\cite{superk}, SNO
CC~\cite{snoccnc}, and SNO NC~\cite{snoccnc} experiments. In this
simplified analysis, we ignore all spectral energy data and all
measurements of the zenith-angle (or day-night) dependence. We let the
total $^8$B neutrino flux be a free variable in the standard
rates-only analysis. For simplicity, we consider pure active, or pure
sterile, neutrino oscillations. Therefore, the rates-only analysis has
three degrees of freedom ($\Delta m^2, \, \tan^2 \theta$, and the
$^8$B neutrino flux) and five measured rates.

The best-fit solution lies in the vacuum region. The oscillation
parameters of the best fit solution are: ${\Delta m^2 = 8 \times
10^{-11} {\rm \, eV^2}, \, \tan^2 \theta = 0.30 (3.35)}$ with
$\chi^2_{\rm min} = 2.2$. The values of $\chi^2_{\rm min}$ for the
other well known solutions are: SMA (3.1) , LMA (3.8), and LOW (8.6),
and pure sterile (23.1). All four solutions for active neutrinos, VAC,
SMA, LMA, and LOW are allowed at $3\sigma$ in the rates-only
analysis. If only rates are considered, the VAC, SMA, and LMA
solutions are essentially equivalent.

 We have done several tests to determine the robustness of this
 conclusion. The vacuum solutions always appear to be slightly
 preferred in the rates-only analyses, but the relative order of
 SMA-LMA can be reversed by slightly changes (such as using the
 original SNO CC data~\cite{sno01} with a 6.75 MeV threshold rather
 than the more recent data~\cite{snoccnc} with a 5.0 MeV threshold).

 This analysis illustrates the fact that the measurement of the
 essentially undistorted recoil electron energy spectrum by
 Super-Kamiokande~\cite{superk}, which was confirmed by
 SNO~\cite{snoccnc}, was required in order to establish, prior to the
 KamLAND measurements, that LMA was the favored solution. Previous
 rates-only analyses have established the critical importance of the
 Super-Kamiokande spectrum analysis~\cite{bks,postsnocc}.  However,
 this is the first calculation that we know about in which all five of
 the currently available rate experiments have been included.

\subsection{KamLAND plus Solar Analysis}
\label{subsec:kamlandplussolar}

It is convenient and conventional to discuss the experimentally
determined $^8$B neutrino flux in units of the flux predicted by the
standard solar model. Let $f_{\rm B,\, total}$ be the ratio of the
total $^8$B solar neutrino flux to the flux predicted by the BP00
standard solar model, i.e.,
\begin{equation}
f_{\rm B,\, total} ~\equiv~ \frac{\phi(^8{\rm B})}{~\phi(^8{\rm
B})_{\rm BP00}}\,,
\label{eq:fbtotaldefinition}
\end{equation}
where $\phi(^8{\rm B})_{\rm BP00}~\equiv~5.05\times 10^6 \,{\rm cm^{-2}s^{-1}}$~\cite{bp00}.

We allow for the possibility that the $^8$B neutrino flux that arrives at Earth contains a significant
component of sterile neutrinos which we parameterize in terms of the active-sterile admixture parameter
$\sin^2\eta$ (See Sec.~\ref{subsubsec:sterile}).

We calculate the global $\chi^2$ by fitting to all the available data,
solar plus reactor.  Formally, the global $\chi^2$ can be written in
the form

\begin{equation}
\chi^2_{\rm global} ~=~ \chi^2_{\rm solar}(\Delta m^2,\tan^2\theta, f_{\rm B,\,total}, \sin^2\eta) ~+~
\chi^2_{\rm KamLAND}(\Delta m^2,\tan^2\theta) \,. \label{eq:chisquaredpowerful}
\end{equation}

There are four free parameters in $\chi^2_{\rm global}$, $\Delta m^2, \tan^2\theta, f_{\rm B,\,total},$
and $sin^2\eta$, although $\chi^2_{\rm KamLAND}$ depends only on $\Delta m^2,$ and $\tan^2\theta$. The
most computationally intensive task in our analysis of solar plus KamLAND data, computing the
multi-dimensional $\chi^2$ for all of the solar neutrino data, was calculated before the KamLAND
announcement~\cite{kamlandannouncement} and stored. This data set amounted to about 9.4 gigabytes. We do
not include the CHOOZ data~\cite{chooz} explicitly in eq.~\ref{eq:chisquaredpowerful} since the combined
solar data also excluded larger values of $\Delta m^2$.

We have carried out a global analysis~\cite{postsno02} of the combined solar neutrino data (80
measurements), see the discussion in section~\ref{subsubsec:solaranalysis} below, and the KamLAND data
(as described in section~\ref{subsec:kamlandanalysis}). We also consider the possibility that the $^8$B
neutrino flux that arrives at Earth contains a significant component of sterile neutrinos. We describe in
section~\ref{subsubsec:sterile} the aspects of the analysis that are related to sterile neutrinos.

\subsubsection{Solar analysis} \label{subsubsec:solaranalysis}

The details of our calculational methods are described in
ref.~\cite{postsno02}.  We treat the $^8$B solar neutrino flux as a
free parameter, but adopt the predictions of the BP00 solar
model~\cite{bp00} for the other neutrino fluxes and uncertainties.
The free $^8$B strategy adopted in the present paper was, in
ref.~\cite{postsno02}, the preferred analysis procedure among the
three analysis strategies that were used earlier.

We summarize briefly in this subsection the main features of our
analysis of solar neutrino data.

For the solar neutrino analysis, we use 80 data points.  We include
the two measured radiochemical rates, from the
chlorine~\cite{chlorine} and the gallium~\cite{sage02,gallex,gno}
experiments, the 44 zenith-spectral energy bins of the electron
neutrino scattering signal measured by the Super-Kamiokande
collaboration ~\cite{superk}, and the 34 day-night spectral energy
bins measured with the SNO~\cite{snoccnc,snodaynight,snourl}
detector. The average rate for the gallium
experiments~\cite{sage02,gallex,gno} has been updated to $70.8 \pm
4.4$ SNU by including the preliminary GNO data reported at Neutrino
2002~\cite{gno}.

We take account of the BP00~\cite{bp00} predicted fluxes and
uncertainties for all solar neutrino sources except for $^8$B
neutrinos. The Super-Kamiokande~\cite{superk} and
SNO~\cite{snoccnc,snodaynight} experiments can now determine the $^8$B
solar neutrino flux at a level of precision that is comparable with
the precision of the BP00 prediction. Therefore, we treat the total
$^8$B solar neutrino flux as a free parameter to be determined by
experiment and to be compared with solar model predictions.

Following the SNO collaboration~\cite{snodaynight}, we have considered
together (consistent with the way they are measured) the SNO charged
current, neutral current, and electron scattering events. For each
point in neutrino oscillation parameter space, we have computed the 34
day-night spectrum energy bins by convolving the computed survival
probabilities with the appropriate neutrino cross sections, as well as
the experimental energy resolution and sensitivity functions.

We include the errors and their correlations for all of these
observables. We construct an $80 \times 80$ covariance matrix that
includes the effect of correlations between the differ error as off
diagonal elements. The errors and their correlations are described in
in the Appendix of ref.~\cite{sterile} and in section~5.4 of
ref.~\cite{postsno02}

The analysis of the theoretical and experimental uncertainties is
subtle.  Many of the differences in the precise borders of the allowed
oscillation regions that are reported by different groups may be
traced to variations in the ways in which the errors are treated.  We
have performed the calculations accurately, although some of the
refinements we have used are computationally demanding. As described
in the Appendix of ref.~\cite{sterile}, we include the energy
dependencies and the correlations of the chlorine and gallium neutrino
absorption cross sections.  For both the Super-Kamiokande and the SNO
experiments, we take account of the energy dependence and the
correlation of the $^8$B spectral energy shape errors, the absolute
energy scale errors, and the energy resolution errors.

\subsubsection{Sterile component}
\label{subsubsec:sterile}

 We have also considered the possibility that $\nu_e$ oscillates into
a state that is a linear combination of active ($\nu_a$) and sterile
($\nu_s$) neutrino states,
\begin{equation}
\nu_e \to \cos\eta \, \nu_a \,+\, \sin\eta \,\nu_s \, ,
\label{eq:linearcombination}
\end{equation}
where $\eta$ is the parameter that describes the active-sterile
admixture. This admixture arises in the framework of 4-$\nu$ mixing
\cite{four}. The total $^8$B neutrino flux can be written
\begin{equation}
\phi({\rm ^8B})_{\rm total} ~=~
\phi(\nu_e)~+~\phi(\nu_a)~+~\phi(\nu_{s}), \label{eq:fluxes}
\end{equation}
where $\phi(\nu_{s})~=~\tan^2 \eta \times \phi(\nu_a)$.

The solar neutrino data analysis depends on four parameters: the oscillation parameters $\Delta m^2$ and
$\tan^2\theta$, the total $^8$B neutrino flux $f_{\rm B,\, total}$ [see
eq.~(\ref{eq:fbtotaldefinition})], and the sterile component characterized by $\sin^2\eta$. We can also
characterize the sterile contribution to the total flux in terms of $f_{\rm B,\,sterile}$, which is
defined as
\begin{equation}
f_{\rm B,\, sterile} ~\equiv~ \frac{~\phi(^8{\rm B,\, sterile})}{\phi(^8{\rm B})_{\rm BP00}}\,.
\label{eq:fbsteriledefinition}
\end{equation}
We obtain from eqs.~(\ref{eq:fluxes}) and~(\ref{eq:linearcombination}) the relation
\begin{equation}
f_{\rm B,\,sterile}=f_{\rm B,\, total} \sin^2\eta \left(1-P_{ee}(E,\Delta m^2, \tan^2\theta,
\sin^2\eta)\right).
\end{equation} Thus for a given set of parameters $\Delta
m^2$, $\tan^2\theta$, $f_{\rm B,\, total}$ and $\sin^2\eta$, the inferred sterile flux parameter $f_{\rm
B,\,sterile}$ is an energy dependent, and time dependent, function. However, the energy and time
dependences are mild since, for the LMA solution, $P_{ee}$ is nearly constant for the $^8$B neutrino
energies that are accessible to direct measurement.

In this paper, we account for the mild energy and time dependences by defining $f_{\rm B,\,sterile}$ as
the average sterile flux parameter inferred for the SNO experiment, i.e.,
\begin{equation}
f_{\rm B,\,sterile}=f_{\rm B,\, total} \sin^2\eta \left(1-\langle
P_{ee}(E,\Delta m^2, \tan^2\theta, \sin^2\eta)\rangle_ {\rm
SNO}\right)
\end{equation}
where $\langle P_{ee}(E,\Delta m^2, \tan^2 \theta, \sin^2\eta)\rangle_
{\rm SNO}$ is
\begin{equation}
\langle P_{ee} (\Delta m^2,\tan^2\theta)\rangle_{\rm SNO}~=~
\frac{\int\! dE_\nu\, \phi^{\rm SSM}({\rm ^8B},\,E_\nu)
\sigma_e(E_\nu) P_{ee} (E_\nu,\Delta m^2,\tan^2 \theta)} {R^{\rm
SSM}_{\rm SNO}}\,.
\label{eq:defnpee}
\end{equation}
Here $P_{ee}$ is the average survival probability and $R^{\rm
SSM}_{\rm SNO}$ is the predicted SSM~\cite{bp00} rate for the CC SNO
measurement~\cite{snoccnc,sno01} in the absence of neutrino
oscillations,
\begin{equation}
R^{\rm SSM}_{\rm SNO} = \int\! dE_\nu\, \phi^{\rm SSM} ({\rm ^8B},\,E_\nu) \sigma_e(E_\nu).
\label{eq:defnrssmsno}
\end{equation}
The cross section $\sigma_{e}(E_\nu)$~\cite{deuteriumcs,deuteriumcsfieldtheory} is the weighted average
cross-section for the CC reaction, $\nu_e + {^2H} ~\rightarrow~ e^+ + n + p$, for the neutrino energy
$E_\nu$ . The average includes the experimental energy resolution function, ${\rm Res} (T,\,T^\prime)$,
where $T (T^\prime)$ is the measured (true) recoil kinetic energy of the electron. Thus
\begin{equation}
\sigma_{e}(E_\nu)=\int_{T_{\rm min}}^{T_{\rm max}}\!dT \int_0^{{T_{\rm
max}}^\prime(E_\nu)}\!dT^\prime\,{\rm Res}(T,\,T^\prime)\,
\frac{d\sigma_{e}(E_\nu,\,T^\prime)}{dT^\prime}\
. \label{eq:sigmaccsno}
\end{equation}
The lower limit, $T_{\rm min}$, in the integral in
eq.~(\ref{eq:sigmaccsno}) is taken here to be the threshold, 5 MeV,
used by the SNO Collaboration in ref.~\cite{snoccnc}. The calculated
value for the CC rate is not sensitive to the assumed value of $T_{\rm
max}$, as long as $T_{\rm max} \geq 17$\,MeV.

We construct $\chi^2_{\rm solar}(\Delta m^2,\tan^2\theta, f_{\rm
B,\,total}, \sin^2\eta)$ in four--dimensional neutrino
parameter space. The results shown in figs.~\ref{fig:before}
and~\ref{fig:after} are the two dimensional allowed regions in $\Delta
m^2, \tan^2 \theta$ that are obtained after marginalizing $\chi^2_{\rm
solar}$ and $\chi^2_{\rm global}$ [defined in
eq.~(\ref{eq:chisquaredpowerful})] over $f_{\rm B,\,total}$ and
$\sin^2\eta$. The allowed domain of the oscillation
parameters at a given CL is defined from the condition
\begin{equation}
\chi^2_{\rm marginalized} (\Delta m^2,\tan^2\theta)~\leq~
\chi^2_{min}+\Delta \chi^2(2,\rm CL)
\label{eq:marginalized}
\end{equation}
where $\chi^2_{\rm min}=81.2$ is the global minimum for the solar
plus KamLAND analysis (for $89=80+13-4$ dof)
in the four-dimensional neutrino parameter space and $\chi^2_{\rm
marginalized} (\Delta m^2,\tan^2\theta)$ is obtained by minimizing
with respect to $f_{\rm B,\,total}$ and $\sin^2\eta$ either
$\chi^2_{\rm solar}$ or $\chi^2_{\rm global}$ for each value of
$\Delta m^2$ and $\tan^2\theta$.

Conversely, in section~\ref{sec:total} (and~\ref{sec:sterile}) we determine $f_{\rm B,\,total}$ (and the
sterile contribution) by minimizing $\chi^2_{\rm global}$ with respect to $\Delta m^2, \tan^2 \theta,$
and $\sin^2\eta$ (or $f_{\rm B,\,total}$). The allowed ranges for each of these parameters at a given CL
are obtained from the corresponding condition for 1 dof. For instance, the allowed range of  $f_{\rm
B,\,total}$ is obtained by applying condition eq.~(\ref{eq:condit1dof}), and similarly for $\sin^2\eta$ 
(or $f_{\rm B,\,sterile}$).

\section{The total $^8$B solar neutrino flux}
\label{sec:total}

One can determine the allowed range of the total $^8$B neutrino flux
using the results of the KamLAND reactor neutrino experiment and the
results of solar neutrino experiments~ref.~\cite{sterile}. The
conceptually simplest approach is to use just the KamLAND
determination of the neutrino oscillation parameters
(cf.~figure~\ref{fig:kamlandonly}) together with the measured CC event
rate in the SNO~\cite{snoccnc} experiment to infer the total $^8$B
solar neutrino flux.  However, this KamLAND-only approach, while very
intuitive, does not make use of the greatly reduced allowed regions
obtained when solar and KamLAND data are combined
(cf.~figure~\ref{fig:before} and figure~\ref{fig:after}). With the
present accuracy of the KamLAND data set, the KamLAND-only
determination yields an imprecise value for the total $^8$B neutrino
flux ($1\sigma$ uncertainty $\sim 30$\%).

Therefore, we present in this section the more accurate total $^8$B
solar neutrino flux that is derived from the four--dimensional global
analysis of solar plus KamLAND neutrino data (see definition of
$\chi^2_{\rm global}$ in eq.~(\ref{eq:chisquaredpowerful}) and
discussion in section~\ref{subsec:kamlandplussolar}). Our results for
$f_{\rm B,\, total}$ are summarized in columns three and four of
table~\ref{tab:bestfits}.

\TABLE[!ht]{ \caption{{\bf The total and the sterile $^8$B solar neutrino flux.} For the two currently
allowed LMA islands in the space of neutrino oscillation parameters that are shown in
figure~\ref{fig:after}, the table presents the best-fit values for the total $^8$B solar neutrino flux
$f_{\rm B}({\rm total})$ and the sterile component of the $^8$B neutrino flux, $f_{\rm B}({\rm
sterile})$.  The values are given in units of the predicted standard solar model $^8$B neutrino
flux~\cite{bp00}. We also present the associated 3$\sigma$ allowed ranges for the global solar plus
KamLAND analysis. The different allowed regions are labeled by the best-fit point in oscillation
parameter space. \label{tab:bestfits}}
\begin{tabular}{ccccccc}
\hline\noalign{\smallskip} $\Delta m^2$&$\tan^2 \theta$& $\chi^2$ & $f_{\rm B,total}$ & 3$\sigma$ $f_{\rm
B,total}$ & $f_{\rm B,sterile}$ &
3$\sigma$ $f_{\rm B,sterile}$\\
\noalign{\smallskip}\hline\noalign{\smallskip} $7.1\times10^{-5}$ & $4.5\times10^{-1}$ &81.2
& 1.00 & [0.81,1.36] & 0.0 & [0.0,0.45]\\
\noalign{\medskip} $1.5\times10^{-4}$ & $4.3\times10^{-1}$ &87.3 & 0.88 & [0.83,0.92] & 0.0 &
[0.0,0.08]\\ \noalign{\smallskip}\hline
\end{tabular} }

In determining $f_{\rm B,\,total}$, we can use all of the available
solar neutrino data as well as the KamLAND measurements. This method
is much more computationally demanding than the more intuitive
KamLAND-only approach since in the global approach one includes, in
addition to the KamLAND data, all of the 80 solar measurements and
their associated (sometimes correlated) uncertainties. Formally, we
calculate (as in section~\ref{sec:global}) a global $\chi^2$, see the
definition of $\chi^2_{\rm global}$ defined in
eq.~(\ref{eq:chisquaredpowerful}).

We obtain the allowed range of $f_{\rm B,\,total}$, the total $^8$B
solar neutrino flux in units of the BP00 predicted flux [see
definition in eq.~(\ref{eq:fbtotaldefinition})] by marginalizing
$\chi^2_{\rm global}$ with respect to $\Delta m^2, \tan^2\theta$, and
$\sin^2\eta$. To be explicit, we minimize
$\chi^2_{\rm global}$ for each value of $f_{\rm B,\,total}$ with
respect to $\Delta m^2, \tan^2\theta$, and $\sin^2\eta$. The
corresponding range of $f_{\rm B,\,total}$ at a given CL is obtained
from the condition
\begin{equation}
\chi^2_{\rm marginalized} (f_{\rm B,\,total})~\leq~ \chi^2_{min}+\Delta \chi^2(1,\rm CL)\,.
\label{eq:condit1dof}
\end{equation}

 We find, after extensive numerical explorations of the
 four-dimensional space of $\chi^2_{\rm global}$, that the current
 best empirical estimate for $f_{\rm B,\,total}$ is
\begin{equation}
f_{\rm B, total} ~=~1.00(1\pm 0.06)~(1\sigma)
\,. \label{eq:bestfbsigma}
\end{equation}
The result given in eq.~(\ref{eq:bestfbsigma}) agrees with the
standard solar model prediction~\cite{bp00}.

In table~\ref{tab:bestfits}, we show the ranges for $f_{\rm B,\,
total}$ for within each of the LMA islands in figure~\ref{fig:after}.

\section{The sterile component of the $^8$B solar neutrino flux}
\label{sec:sterile}

The sterile component of the $^8$B solar neutrino flux can be determined directly by subtracting the
active $^8$B neutrino flux, which is given by the SNO NC measurement~\cite{snoccnc}, from the total $^8$B
neutrino flux determined in section~\ref{sec:total} (cf.~refs.~\cite{sterile,holanda}). This subtraction
can be made in a model independent fashion using the oscillation parameters obtained from the analysis of
the KamLAND data alone. However, this direct method, while conceptually simple, yields an imprecise
determination of the sterile component of the $^8$B neutrino flux ($1\sigma$ uncertainty of order 18\%).

The more powerful approach is to include all of the solar plus KamLAND measurements
(cf.~section~\ref{sec:total}). In this approach, we first evaluate the global $\chi^2$, $\chi^2_{\rm
global}(\Delta m^2, \tan^2 \theta, f_{\rm B,\,total},\sin^2\eta )$, which is defined in
eq.~(\ref{eq:chisquaredpowerful}). Not only are all of the relevant data used in calculating $\chi^2_{\rm
global}$, but also this method properly includes the correlated uncertainties. Thus calculating
$\chi^2_{\rm global}$ is the preferred statistical method as well as the more powerful (but
computationally intensive) method.

 As discussed in refs.~\cite{sterile,bargersterile}, the allowed
sterile component in the $^8$B neutrino flux is strongly correlated with the allowed departure from the
standard solar model prediction of the total $^8$B neutrino flux. We illustrate this correlation in
figure~\ref{fig:ftotal_eta}.

We show in figure~\ref{fig:ftotal_eta} the allowed regions in the $f_{\rm B,\,total}-\sin^2\eta$
parameter space after marginalizing with respect to $\Delta m^2$ and $\tan^2 \theta$. For the sake of
clarity, we show separately the results of marginalizing for both of the LMA islands shown in
figure~\ref{fig:after}.

\FIGURE[!ht]{\centerline{\psfig{figure=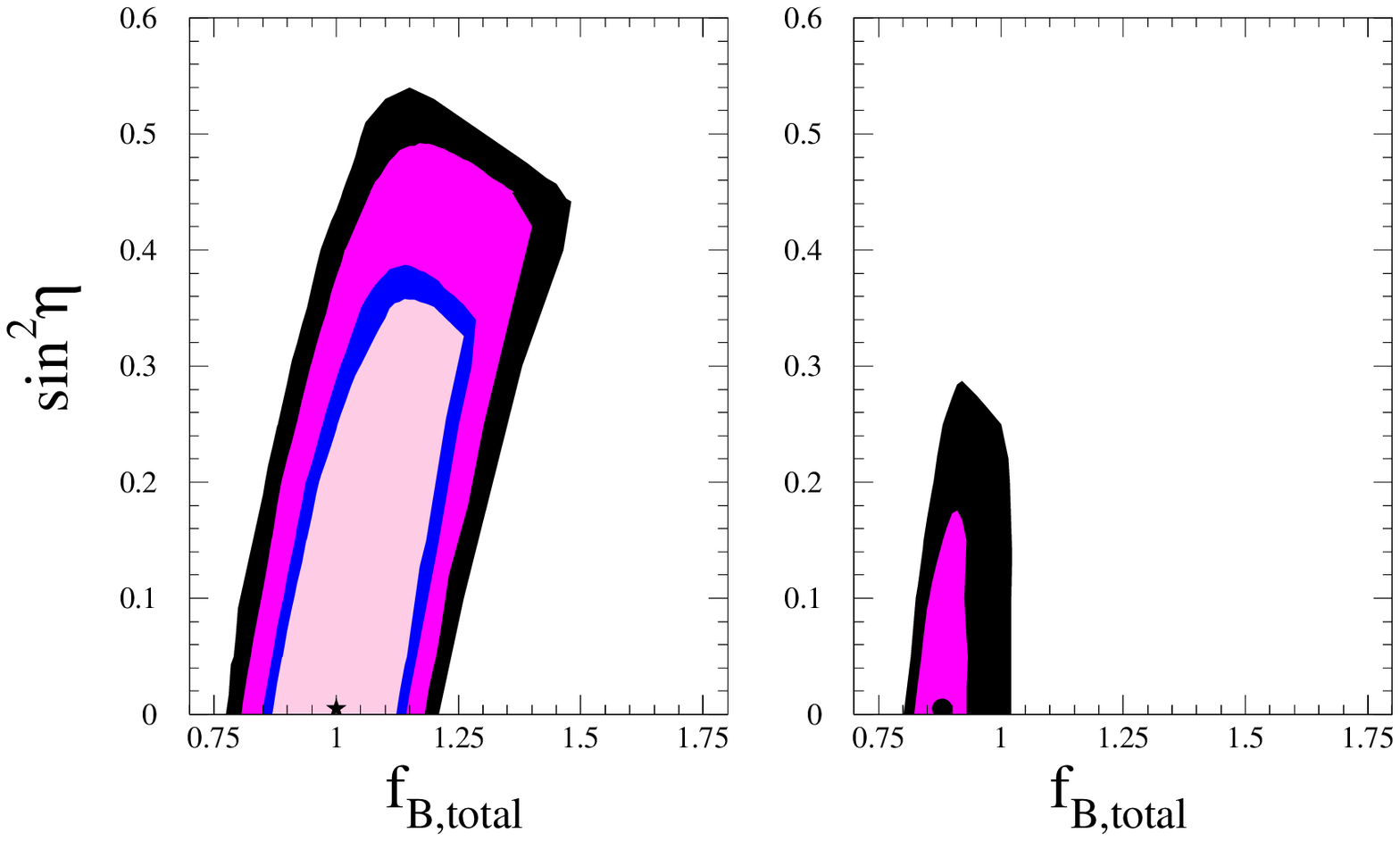,width=4.5in}} \caption{{\bf Limit on active-sterile
admixture.} The figure shows the active-sterile admixture $\sin^2\eta$ as a function of $f_{\rm B,\,
total}$, the total $^8$B flux relative to the BP00 predicted standard $^8$B neutrino flux . The contours
shown in figure~\ref{fig:ftotal_eta} are $90$\%, $95$\%, $99$\%, and $99.73$\% ($3\sigma$)confidence
limits. \label{fig:ftotal_eta}}}

In figure~\ref{fig:ftotal_fsterile}, we display the corresponding allowed regions in terms of $f_{\rm
B,\,total}-f_{\rm B,\, sterile}$ (after marginalizing with respect to $\Delta m^2$ and $\tan^2 \theta$).

\FIGURE[!ht]{\centerline{\psfig{figure=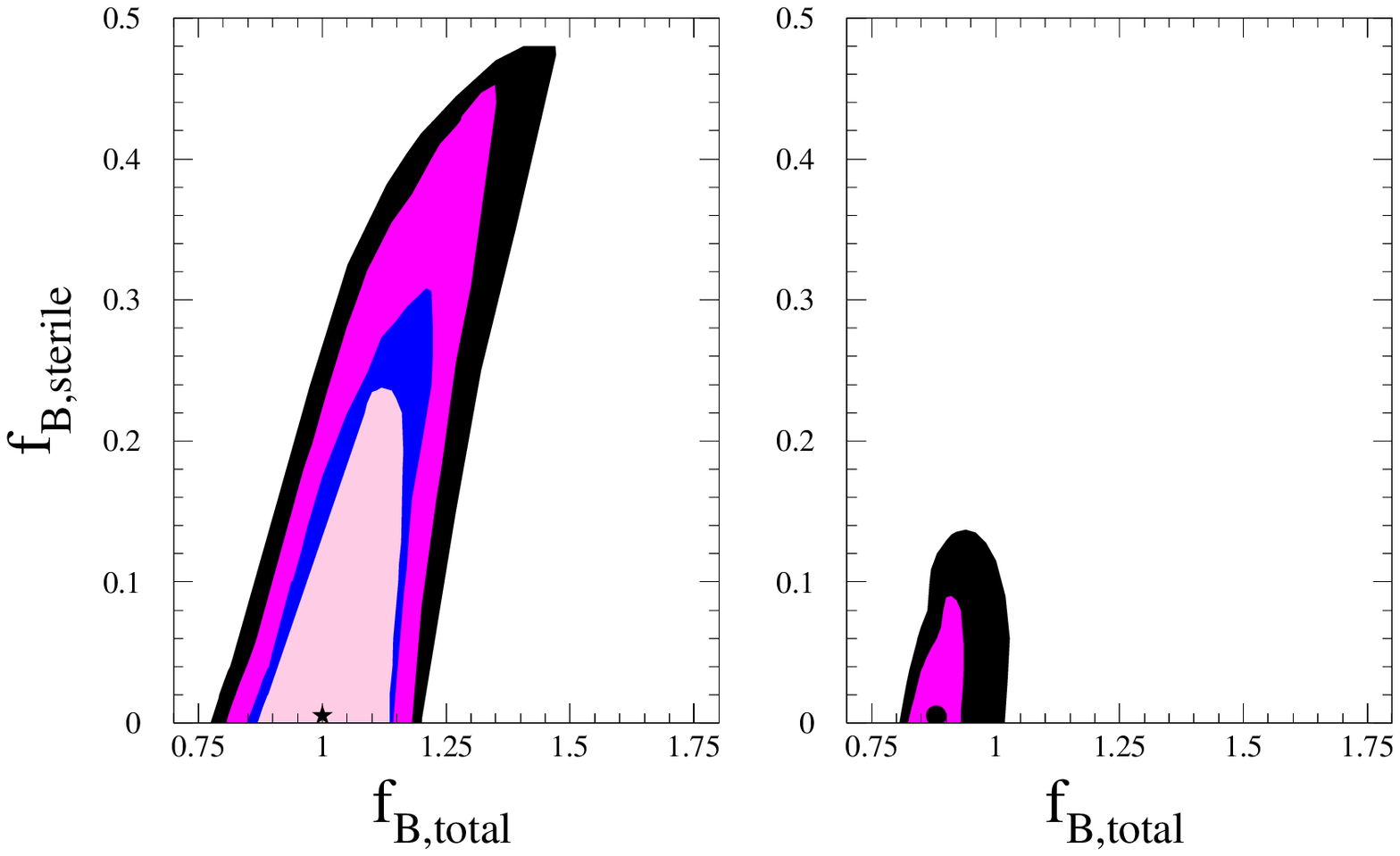,width=4.5in}} \caption{{\bf The Sterile-Total
flux correlation .} The figure shows the sterile component, $f_{\rm B,\,
sterile}$ as a function of $f_{\rm B,\, total}$, the total $^8$B flux, both relative to the BP00
predicted standard $^8$B neutrino flux . The contours shown in figure~\ref{fig:ftotal_fsterile} are
$90$\%, $95$\%, $99$\%, and $99.73$\% ($3\sigma$)confidence limits. \label{fig:ftotal_fsterile}} }

Finally,  marginalizing with respect to $\Delta m^2, \tan^2 \theta,{~\rm and} f_{\rm B,\,total}$, we
obtain the allowed range of the active-sterile admixture
\begin{equation}
\sin^2\eta\leq 0.13~(0.52)\; , \label{eq:etaklandplussolar}
\end{equation}
at 1$\sigma$ (3$\sigma$).

Correspondingly for $f_{\rm B,\, sterile}$, we show in columns five and six of table~\ref{tab:bestfits}
the best-fit values and the 3$\sigma$ ranges for the local minima of the two currently allowed LMA
islands. The $1\sigma$  range from this global analysis is
\begin{equation}
f_{\rm B,\, sterile}~=~0.00^{+0.09}_{-0.0}\,. \label{eq:fbsterileklandplussolar}
\end{equation}
We use eq.~(\ref{eq:fbsterileklandplussolar}) as our best-estimate of the sterile contribution to the
observed $^8$B solar neutrino flux.

Is the limit on $f_{\rm B,\, sterile}$ given in eq.~(\ref{eq:fbsterileklandplussolar}) (or on
$\sin^2\eta$ in eq.~(\ref{eq:etaklandplussolar}) changed significantly if a more general mixing of
sterile neutrinos is considered? We have investigated numerically cases in which a second $\Delta m^2$
affects solar neutrino oscillations. We have adopted as a paradigm the model discussed in
ref.~\cite{holanda}, in which there is a third neutrino mass scale, $\Delta m^2\sim 10^{-6} $eV$^2$,
related to a sterile neutrino. We find that, within the accuracy indicated by
eq.~(\ref{eq:fbsterileklandplussolar}), our best-estimate of the sterile contribution (and the associated
uncertainty) of the $^8$B solar neutrino flux is not affected by the presence of this lower mass scale.
The main effect of this low-mass sterile neutrino scenario is that the sterile neutrinos add an extra
energy dependence to the survival probability and to the sterile fraction. So the limit given in
eq.~(\ref{eq:fbsterileklandplussolar}) for the $^8$B solar neutrino flux will not apply to the sterile
component of the lower energy $p-p$ and $^7$Be solar neutrino fluxes. The sterile component for the $p-p$
and $^7$Be neutrino fluxes must be determined by a combination of CC and $\nu-e$ scattering experiments
at low energies ($< 1$ MeV).

\section{Summary}
\label{sec:summary}

The allowed oscillation regions that are determined by including the initial KamLAND
results~\cite{kamlandannouncement} together with the previously known solar neutrino results (see
figure~\ref{fig:after}) populate 6 orders of magnitude less area in neutrino oscillation parameter space
($\Delta m^2, \tan^2\theta$) than the allowed oscillation regions that are determined with just solar
neutrino data (see figure~\ref{fig:before}). The LMA solution is now the only allowed solution at
$3\sigma$. The LOW, Vacuum, and SMA solutions are disfavored at, respectively, $4.7\sigma$, $4.8\sigma$,
and $6.1\sigma$. All alternative explanations to the solar neutrino problem that do not also predict a
deficit in KamLAND are now excluded at 3.6$\sigma$. Combining solar neutrino data with the KamLAND
measurements,  non-standard solar model explanations of the solar neutrino problem that do not include
new particle physics (e.g., neutrino oscillations) are excluded at the $7.9\sigma$ CL. Maximal mixing,
$\tan^2 \theta = 1$, is disfavored at $3.1\sigma$. The results in this paragraph are described in
sec~\ref{subsubsec:globalallowedregions}.

At the present level of accuracy, matter effects are not important in
determining the allowed oscillation regions permitted by the KamLAND
experiment. However, when the precision of the experiment is improved,
matter effects must be included (see
sec.~\ref{subsec:mattereffects}). Within the currently allowed
oscillation regions, the maximum change is 4\% in the calculated event
rate for the KamLAND experiment for a specific value of $\Delta m^2$
and $\tan^2 \theta$.

The LMA prediction of the $\nu-e$ scattering rate for $^7$Be solar neutrinos is precise, $0.64\pm 0.03$
within the $1\sigma$ (2 dof)currently allowed oscillation region (see
section~\ref{subsubsec:futureexperiments}), in units of the rate predicted by the standard solar model
and assuming no flavor conversion. The estimated uncertainty in the predicted standard solar model $^7$Be
solar neutrino flux, $\pm 10$\%~\cite{bp00}, is much larger than the current uncertainty in the predicted
scattering rate due to neutrino oscillation parameters. Thus BOREXINO and KamLAND can provide a crucial
test of solar model predictions by measuring accurately the $^7$Be $\nu-e$ scattering rate.

Table~\ref{tab:predictions} of section~\ref{subsubsec:futureexperiments} summarizes the predicted rates
for $\nu-e$ scattering by $p-p$ as well as $^7$Be ( and $^8$B) neutrinos. Within the $1\sigma$ (2 dof)
allowed region, the predicted $p-p$ scattering rate is $\sim \pm 2$\% due to imperfectly known
oscillation parameters, which is about twice the currently estimated uncertainty in the solar model
prediction of the $p-p$ neutrino flux. If our current ideas regarding the accuracy of the solar model
predictions are valid and if the popular neutrino oscillation framework is correct, then a measurement of
the $p-p$ solar neutrino will provide a precise determination of the mixing angle $\theta$.

It will be difficult with the existing SNO detector~\cite{snoccnc,snodaynight} to measure the predicted
SNO CC day-night asymmetry: $A_{\rm N-D}({\rm SNO~CC}) ~=~ 3.3^{+0.7}_{-0.6} \% (~1\sigma )$ (see
eq.~\ref{eq:CCdaynight}).

The event rate measured at KamLAND,
(eq.~\ref{eq:kamlandrate}) lies within the 1-sigma pre-KamLAND
expectations  based on purely solar neutrino experiments
(eq.~\ref{eq:kamlandratepred}). This excellent agreement indicates that the
main features of solar neutrinos are beginning to be well understood.

The most accurate method to determine the total $^8$B flux is to
combine together all of the solar as well as KamLAND experimental
data. Using this computationally intensive method, we find (see
section~\ref{sec:total}) that the total $^8$B solar neutrino flux is
$1.00\left(1\pm 0.06\right)$ in units of the standard solar model
predicted flux~\cite{bp00}.

The sterile component of the $^8$B neutrino flux can be determined similarly by using all of the
available solar and KamLAND data. The allowed range of the active-sterile admixture is $\sin^2\eta\leq
0.13$. Correspondingly, our best estimate for the sterile component of the $^8$B solar neutrino flux is
$0.00^{+0.09}_{-0.00}$ in units of the standard solar model predicted flux (see
section~\ref{sec:sterile}).

Our results for the total $^8$B solar neutrino flux, and for the
sterile component of this flux, are summarized in
table~\ref{tab:bestfits}. The present determination of the total $^8$B
solar neutrino flux reduces the $1\sigma$ experimental uncertainty in this
quantity by a factor of two with respect to the previous most accurate
determination~\cite{sterile}.  The experimental bounds on the total
$^8$B neutrino flux are now more than a factor of two smaller that the
uncertainties in the predicted standard solar model flux~\cite{bp00}.
This is the first ocassion on which the experimental errors on a solar
neutrino flux are smaller than the solar model uncertainties. The
uncertainty in the sterile neutrino component of the $^8$B neutrino flux
represents a factor of two improvement over the previous most
stringent bound.

We conclude with an important methodological question. Do the results
given in this paper depend in a significant way upon details of the
analysis of the KamLAND data? This is a natural question to ask since
the KamLAND results represent a crucial but low statistics addition to
the solar neutrino data.  Fortunately, different analysis strategies
lead to essentially the same quantitative inferences regarding solar
neutrino properties and predictions.

We have studied the robustness of our quantitative conclusions to
different plausible variations in the statistical analysis of the
KamLAND data. Most of these variations made no discernible difference
in the final answers that were obtained from the global solar plus
KamLAND analysis. For quantities such as the total $^8$B neutrino
flux, the sterile component of the $^8$B neutrino flux, and the
predictions for observables in future experiment, there were no
significant differences.  For specificity, we report here the results
of a $\chi^2$ analysis of the KamLAND data which led to the largest
changes of any of the alternative analysis strategies we adopted. In
this $\chi^2$ analysis, we used the first 5 energy bins above 2.6 MeV
visible energy as presented by the KamLAND collaboration but collected
all the events above the fifth bin ($E > 4.725$ MeV) into a single bin
(or into two bins) in order to have enough events in each bin to
assume Gaussian errors. We found that the allowed regions of the
KamLAND-only analysis are slightly modified, but in ways that would be
noticed only by the trained and expert eye (not to the casual
reader). However, the effect of using a standard $\chi^2$ with
Gaussian errors is even less important for determining the global,
solar plus KamLAND, allowed regions. In particular, this alternative
analysis changes $f_{\rm B,\, total}$ by $\pm 0.01$ ($1\sigma$) and
$f_{\rm B,\, sterile}$ by $\pm 0.005$ ($1\sigma$).  The maximum change
in the range of predicted solar neutrino observables in
table~\ref{tab:predictions} occurs for the $R(^8{\rm B})$ and amounts
to $2$\% at 1$\sigma$ in the largest allowed value. For all other
observables, the changes are less than $2$\% in the $3\sigma$ range.

We are grateful E. Akmedov and M. Smy for helpful comments and to G. Feldman for asking the important
question of how large are the matter effects in the KamLAND experiment. JNB and CPG acknowledge support
from NSF grant No. PHY0070928. MCG-G is supported by the European Union Marie-Curie fellowship
HPMF-CT-2000-00516. This work was also supported by the Spanish DGICYT under grants PB98-0693 and by
Generalitat Valenciana CTIDIB/2002/24.


\clearpage
\begin{thebibliography}{999}
\bibitem{kamlandannouncement}K. Eguchi et al., the KamLAND collaboration,
{\it First results from KamLAND: Evidence for reactor anti-neutrino
disappearance}, hep-ex/0212021.
\bibitem{sterile}J.N. Bahcall, M.C. Gonzalez-Garcia and C. Pe\~na-Garay, {\it If
sterile neutrinos exist, how can one determine the total solar
neutrino fluxes?},
{\it Phys. Rev.} {\bf C 66} (2002) 035802 [hep-ph/0204194].
\bibitem{friedland}A. Friedland and A. Gruzinov, {\it Has
Super-Kamiokande observed antineutrinos from the sun?}, hep-ph/0202095; P. Aliani, V. Antonelli, M.
Picariello and E. Torrente-Lujan, {\it KamLAND, solar antineutrinos and their magnetic moment},
hep-ph/0208089; E. Kh. Akhmedov and J. Pulido, {\it Solar neutrino oscillations and bounds on neutrino
magnetic moment and solar magnetic field}, hep-ph/0209192.
\bibitem{cptconserv}J.N. Bahcall, V. Barger and D. Marfatia, {\it How
accurately can one test CPT conservation with reactor and solar
neutrino experiments?}, {\it Phys. Lett.} {\bf B 534} (2002) 120.
\bibitem{beacom}J.F. Beacom and N.F. Bell, {\it Do solar neutrinos
decay?}, {\it Phys. Rev.} {\bf D 65} (2002) 113009.
\bibitem{shrock}I. Mocioiu and R. Shrock, {\it Neutrino oscillations
with two $\Delta m^2$ scales}, \jhep{11}{2001}{050}.
\bibitem{aliani-1}P. Aliani, V. Antonelli, M. Picariello and
E. Torrente-Lujan, {\it KamLAND potentiality on the determination of neutrino mixing parameters in the
post SNO-NC era}, hep-ph/0207348.
\bibitem{grimus}W. Grimus, M. Maltoni, T. Schwetz, M.A. T\'ortola and
J.W.F. Valle, {\it Constraining Majorana neutrino electromagnetic properties from the LMA-MSW solution of
the solar neutrino problem}, hep-ph/0208132.
\bibitem{holanda}P.C. de Holanda and A.Yu. Smirnov, {\it Searches for
sterile component with solar neutrinos and KamLAND}, hep-ph/0211264.
\bibitem{bandyo}A. Bandyopadhyay, S. Choubey, R. Gandhi, S. Goswami
and D.P. Roy, {\it Testing the solar LMA region with KamLAND data}, hep-ph/0211266.
\bibitem{cnopaper}J.N. Bahcall, M.C. Gonzalez-Garcia and
C. Pe\~na-Garay, {\it Does the sun shine by p-p or CNO reactions?}, astro-ph/0212331.
 \bibitem{postsno02}J.N. Bahcall, M.C. Gonzalez-Garcia and
C. Pe\~na-Garay, {\it Before and after: how has the SNO NC measurement
changed things?},
\jhep{07}{2002}{054} [hep-ph/0204314].
\bibitem{chlorine}B.T. Cleveland et al., {\it Measurement of the solar
electron neutrino flux with the Homestake chlorine detector},
{\it Astrophys. J.}
{\bf 496} (1998) 505.
\bibitem{sage02}
SAGE collaboration, J.N. Abdurashitov et al., {\it Measurement of the solar neutrino capture rate by the
Russian-American gallium solar neutrino experiment during one half of the 22-year cycle of solar
activity}, {\it J. Exp. Theor. Phys.} {\bf 95} (2002) 181 [astro-ph/0204245].
\bibitem{gallex}GALLEX collaboration, W. Hampel et al., {\it GALLEX
solar neutrino observations: results for GALLEX IV}, {\it Phys. Lett.} {\bf B 447} (1999) 127.
\bibitem{gno}T. Kirsten, {\it Progress in GNO}, talk at the XXth International Conference on Neutrino Physics
and Astrophysics (NU2002), Munich, (May 25-30, 2002); GNO collaboration, M. Altmann et al., {\it GNO
solar neutrino observations: results for GNO I}, {\it Phys. Lett.} {\bf B 490} (2000) 16.
\bibitem{superk}Super-Kamiokande Collaboration,
S.~Fukuda  et al., {\it Solar ${\rm ^8B}$ and hep neutrino
measurements from 1258 days of Super-Kamiokande data},
{\it Phys. Rev. Lett.} {\bf 86} (2001) 5651.
\bibitem{snoccnc}SNO collaboration, Q.R. Ahmad et al., {\it Direct evidence for neutrino flavor transformation from
neutral-current interactions in the Sudbury Neutrino Observatory}, {\it Phys. Rev. Lett.} {\bf 89} (2002)
011301 [nucl-ex/0204008].
\bibitem{snodaynight}SNO collaboration, Q.R. Ahmad et al., {\it Measurement of day and night neutrino energy spectra
at SNO and constraints on neutrino mixing parameters}, {\it Phys. Rev. Lett.} {\bf 89} (2002) 011302
[nucl-ex/0204009].
\bibitem{bargersterile} V.~Barger, D.~Marfatia, K.~Whisnant, {\it Piecing the Solar Neutrino Puzzle Together at SNO}, {\it
Phys. Lett.} {\bf B 509} (2001) 19.
\bibitem{chooz}M. Apollonio et al., {\it Limits on neutrino
oscillations from the CHOOZ experiment}, {\it Phys. Lett.} {\bf B 466} (1999) 415.
\bibitem{msw}L. Wolfenstein, {\it Neutrino oscillations in matter},
{\it Phys. Rev.} {\bf D 17} (1978) 2369; S.P. Mikheyev and A.Y. Smirnov, {\it Resonance enhancement of
oscillations in matter and solar neutrino spectroscopy}, {\it Sov. Jour. Nucl. Phys.} {\bf 42} (1985)
913.
\bibitem{kamland}KamLAND collaboration, P. Alivisatos et al.,
{\it KamLAND: a liquid scintillator anti-neutrino detector at the Kamioka site}, Stanford-HEP-98-03;
KamLAND collaboration, A. Piepke, {\it KamLAND: a reactor neutrino experiment testing the solar neutrino
anomaly}, in {\it Neutrino 2000}, Proc. of the XIXth International Conference on Neutrino Physics and
Astrophysics, 16--21 June 2000, eds. J. Law, R.W. Ollerhead and J.J. Simpson, \npps{B 91}{2001}{99}.
\bibitem{prekamlandus}
A.~de Gouvea and C.~Pe\~na-Garay, {\it Solving the solar neutrino puzzle with KamLAND and solar data},
{\it Phys.\ Rev.} {\bf D 64} (2001) 113011; M.C.~Gonzalez-Garcia and C.~Pe\~na-Garay, {\it On the effect
of $\theta_{13}$ on the determination of solar oscillation parameters at KamLAND}, {\it Phys. Lett.} {\bf
B 527} (2002) 199.
\bibitem{prekamlandthem}
 R.~Barbieri and A.~Strumia, {\it Non standard analysis of the solar
neutrino anomaly}, \jhep{12}{2000}{016}; V.~Barger,
 D.~Marfatia and B.P.~Wood, {\it Resolving the solar neutrino problem
with KamLAND}, {\it Phys. Lett.} {\bf B 498} (2001) 53; H.~Murayama and A.~Pierce, {\it Energy spectra of
reactor neutrinos at KamLAND}, {\it Phys. Rev.} {\bf D 65} (2002) 013012.
\bibitem{robust}J.N.~Bahcall,
M.C.~Gonzalez-Garcia, and C.~Pe\~na-Garay, {\it Robust signatures of solar neutrino oscillation
solutions}, {\it J. High Energy Phys.} {\bf 04} (2002) 007 [hep-ph/0111150].
\bibitem{pdg}
K.~Hagiwara et al. [Particle Data Group Collaboration], {\it Review Of Particle Physics,} {\it
Phys. Rev.} {\bf D 66} (2002) 010001.
\bibitem{bargernc}V. Barger, D. Marfatia, K. Whisnant and B.P. Wood,
{\it Imprint of SNO neutral current data on the solar neutrino problem}, {\it Phys. Lett.} {\bf B 537}
(2002) 179; P. Creminelli, G. Signorelli and  A. Strumia, {\it Frequentist analyses of solar neutrino
data}, \jhep{05}{2001}{052} [hep-ph/0102234, see addendum 04/22/2002];
 G.L. Fogli, E. Lisi, A. Marrone, D. Montanino and A. Palazzo, {\it
Getting the most from the statistical analysis of solar neutrino oscillations}, {\it Phys. Rev.} {\bf D
66} (2002) 053010; A. Bandyopadhyay, S. Choubey, S. Goswami and D.P. Roy, {\it Implications of the first
neutral current data from SNO for solar neutrino oscillation}, {\it Phys. Lett.} {\bf B 540} (2002) 14;
P.C. de Holanda and A.Yu. Smirnov, {\it Solar neutrinos: global analysis with day and night spectra from
SNO}, hep-ph/0205241; Y. Fukuda
 et al., {\it Determination of solar neutrino oscillation parameters
using 1496 days of super-Kamiokande-I data}, {\it Phys. Lett.} {\bf B 539} (2002) 179; A. Strumia, C.
Cattadori, N. Ferrari and F. Vissani, {\it Which solar neutrino data favour the LMA solution?}, {\it
Phys. Lett.} {\bf B 541} (2002) 327; S. Pascoli and S.T. Petcov, {\it The SNO solar neutrino data,
neutrinoless double beta-decay and neutrino mass spectrum}, {\it Phys. Lett.} {\bf B 544} (2002) 239
[hep-ph/0205022]; M. Maltoni, T. Schwetz, M.A. Tortola and J.W.F. Valle, {\it Constraining neutrino
oscillation parameters with current solar and atmospheric data}, hep-ph/0207227.
\bibitem{bp00}J.N. Bahcall, M.H. Pinsonneault and S. Basu, {\it
Solar models: current epoch and time dependences, neutrinos, and helioseismological properties}, {\it
Astrophys. J.} {\bf 555} (2001) 990.
\bibitem{sno01}SNO collaboration, Q.R. Ahmad et al., {\it Measurement of charged
current interactions produced by ${\rm ^8B}$ solar neutrinos at the Sudbury Neutrino Observatory}, {\it
Phys. Rev. Lett.} {\bf 87} (2001) 071301.
\bibitem{luminosity}J. N. Bahcall, {\it The Luminosity Constraint on Solar Neutrino Fluxes}
{\it Phys. Rev.} {\bf C 65} (2002) 025801.
\bibitem{borexino} BOREXINO collaboration, G. Alimonti et al., {\it
Science and technology of BOREXINO: a real time detector for low energy solar neutrinos}, {\it Astropart.
Phys.} {\bf 16} (2002) 205.
\bibitem{obslma}
J.~N.~Bahcall, P.~I.~Krastev and A.~Y.~Smirnov, {\it Is large mixing angle MSW the solution of the
solar neutrino problems?}, {\it Phys.\ Rev.} {\bf D 60} (1999), 093001.
\bibitem{10comm} J.~N.~Bahcall, P.~I.~Krastev and A.~Y.~Smirnov,
{\it SNO: Predictions for ten measurable quantities}, {\it Phys.\ Rev.} {\bf D 62} (2000), 093004.
\bibitem{zenith}
M.~C.~Gonzalez-Garcia, C.~Pena-Garay and A.~Y.~Smirnov, {\it Zenith angle distributions at Super-Kamiokande
and SNO and the solution  of the solar neutrino problem}, {\it Phys.\ Rev.} {\bf D 63} (2001), 113004.
\bibitem{bks}J.N. Bahcall, P.I. Krastev, A.Yu. Smirnov, {\it
Solar neutrinos: global analysis and implications for SNO}, {\it
J. High Energy Phys.} {\bf 05} (2001) 015 [hep-ph/0103179].
\bibitem{postsnocc}J.N.~Bahcall, M.C.~Gonzalez-Garcia, and C.~Pe\~na-Garay, {\it Global analysis of solar neutrino
oscillations including SNO CC measurement}, {\it J. High Energy Phys.} {\bf 08} (2001) 014
[hep-ph/0106258].
\bibitem{snourl}http://www.sno.phy.queensu.ca/. Choose `Latest results from SNO (April 20, 2002).' Click on
`Look at our data base.'
\bibitem{four}D. Dooling, C. Giunti, K. Kang and C.W. Kim, {\it Matter
effects in four-neutrino mixing}, {\it Phys. Rev.} {\bf D 61} (2000)
073011; C. Giunti, M.C. Gonzalez-Garcia and C. Pe\~na-Garay, {\it
Four-neutrino oscillation solutions of the solar neutrino problem},
{\it Phys. rev. } {\bf D 62} (2000) 013005.
\bibitem{deuteriumcs}S. Nakamura, T. Sato, S. Ando, T.-S. Park, F. Myhrer, V. Gudkov and
K. Kubodera, {\it Neutrino-deuteron reactions at solar neutrino energies}, \npa{707}{2002}{561}
[nucl-th/0201062]. See also,  A.~Kurylov, M.J.~Ramsey-Musolf and P.~Vogel, {\it Radiative corrections in
neutrino-deuterium disintegration}, {\it Phys.\ Rev.} {\bf C 65} (2002) 055501;
  J.F.~Beacom and S.J.~Parke, {\it Normalization of the
neutrino-deuteron cross section}, {\it Phys.\ Rev.} {\bf D 64} (2001) 091302.
\bibitem{deuteriumcsfieldtheory}
M. Butler, J.-W. Chen and X. Kong, {\it Neutrino-deuteron scattering in effective field theory at
next-to-next-to-leading order}, {\it Phys. Rev.} {\bf C 63} (2001) 035501.
\end{thebibliography}
\end{document}